\documentclass[12pt]{article}
\usepackage{amsmath}
\usepackage{amssymb}
\begin{document}
\begin{center}
{\bf {\large{Nilpotent Charges in an Interacting Gauge Theory\\ and an 
${\cal N } = 2$ SUSY Quantum Mechanical Model:
(Anti-)Chiral Superfield Approach}}}

\vskip 3.0cm

{\sf  B. Chauhan$^{(a)}$, S. Kumar$^{(a)}$, R. P. Malik$^{(a,b)}$}\\
$^{(a)}$ {\it Physics Department, Institute of Science,}\\
{\it Banaras Hindu University, Varanasi - 221 005, (U.P.), India}\\

\vskip 0.1cm

\vskip 0.1cm

$^{(b)}$ {\it DST Centre for Interdisciplinary Mathematical Sciences,}\\
{\it Institute of Science, Banaras Hindu University, Varanasi - 221 005, India}\\
{\small {\sf {e-mails: bchauhan501@gmail.com; sunil.bhu93@gmail.com; rpmalik1995@gmail.com}}}

\end{center}

\vskip 2cm

\noindent
{\bf Abstract:}
We exploit the power and potential of the (anti-)chiral superfield approach (ACSA)
to Becchi-Rouet-Stora-Tyutin (BRST) formalism to derive the nilpotent (anti-) BRST symmetry 
transformations for any arbitrary $D$-dimensional {\it interacting} non-Abelian 1-form gauge theory where there is an
$SU(N)$ gauge invariant coupling between the gauge field and the Dirac fields. We derive the conserved 
and nilpotent (anti-)BRST charges and establish their nilpotency and absolute anticommutativity properties within
 the framework of ACSA to BRST formalism. The clinching  proof of the absolute anticommutativity property of the conserved
and nilpotent (anti-)BRST charges is a {\it novel} result in view of the fact that we consider,
in our present endeavor, {\it only} the (anti-)chiral super expansions 
 of the superfields that are defined on the $(D, 1)$-dimensional super-submanifolds of the {\it general}
 $(D, 2)$-dimensional supermanifold on which our $D$-dimensional {\it ordinary} interacting non-Abelian 
1-form gauge theory is generalized. To corroborate the novelty of the {\it above} result, we apply the ACSA to an
${\cal N } = 2$ supersymmetric (SUSY) quantum mechanical (QM) model of a harmonic oscillator  and show that the nilpotent
 and conserved ${\cal N } = 2$ super charges of {\it this} system do {\it not} absolutely anticommute. 



\section {Introduction}

\noindent
The abstract mathematical properties (i.e. off-shell nilpotency and absolute anticommutativity)
associated with the Becchi-Rouet-Stora-Tyutin (BRST) and anti-BRST symmetries find their geometrical basis
 within the framework of the usual superfield approach (USFA) to BRST formalism. One of the key concepts behind USFA 
is the idea of horizontality condition (HC) where a particular geometrical quantity  (i.e. an exterior  derivative) plays a decisive role. The central 
outcome  of the HC is the observation that it leads to the derivation of (anti-)BRST symmetries 
for {\it only} the gauge field and associated (anti-)ghost fields of a given (anti-)BRST invariant theory. It does not
shed any light on the derivation of the (anti-)BRST symmetries associated with the {\it matter} fields 
in an {\it interacting} gauge theory. The USFA has been systematically and consistently extended 
 so as to derive the (anti-)BRST symmetry transformations for the gauge, (anti-)ghost 
and {\it matter} fields {\it together}. The extended version of the USFA has been christened  
as the augmented version of superfield approach (AVSA) to BRST formalism [9-12] where, in addition to the 
HC, the gauge invariant restrictions (GIRs) have {\it also} been invoked. The latter are {\it consistent}
with the HC and {\it both} of them complement each-other in a meaningful manner (within 
the framework of AVSA) where we precisely  derive the nilpotent symmetries for {\it all} 
the fields of an interacting gauge theory.

The key feature of the above superfield approaches [1-12] is the fact that 
{\it all} the superfields of the ($D$, 2)-dimensional supermanifold (on which a given
$D$-dimensional {\it ordinary} gauge theory  is  generalized) are expanded 
along {\it all} the possible Grassmannian directions of the supermanifold. This 
supermanifold is parameterized by the superspace coordinates $Z^M = (x^\mu, \theta, \bar\theta)$  
where $x^\mu\;(\mu  = 0, 1,....D-1)$
are the $D$-dimensional bosonic coordinates and a pair of Grassmannian variables satisfy:
$\theta^2 = \bar\theta^2 = 0,\;\theta\bar\theta+ \bar\theta\theta = 0$. The purpose of our
present endeavor is to apply a {\it simpler} version of the  above superfield approaches [1-12] where
{\it only} the (anti-)chiral superfields
are taken into account for the derivation of the {\it proper}
(anti-)BRST symmetry transformations. In a recent set of papers [13-15], we have exploited the 
(anti-)chiral {\it superfield/supervariable} approach (ACSA) to BRST formalism\footnote {The beauty of the (anti-)chiral superfield/supervariable  approach is the observation that we obtain the (anti-)BRST symmetries for {\it all} the fields/variables of the theory from the (anti-)BRST (i.e.
{\it quantum} gauge) invariant restrictions on the (anti-)chiral  superfileds/supervariables.}  to obtain the 
(anti-)BRST symmetries for the $D$-dimensional (non-)interacting {\it Abelian} 1-form and 1$D$ toy models 
 of gauge theories. We have been able to establish that, {\it despite} the
(anti-)chiral superfield/supervariable considerations, the (anti-) BRST charges turn out to be absolutely anticommuting in nature.
This observation is a completely   {\it novel } and surprising result within the framework of ACSA to BRST formalism.
So far, we have {\it not} applied the ACSA to BRST formalism in the case of any arbitrary $D$-dimensional (non-)interacting
 {\it non-Abelian} 1-form gauge theory ({\it with} or {\it without} matter fields).

In our present investigation, we apply, first of all, the ACSA to BRST formalism in the case of a $D$-dimensional {\it interacting} non-Abelian gauge theory
and show that the expressions for the (anti-)BRST charges
for the  {\it interacting} non-Abelian theory are {\it exactly} the same as in the case of non-Abelian 
theory {\it without} matter fields. In other words, as can be seen in the  expressions (see, e.g. Eq. (12) below),
there is a possibility of {\it no} presence of the {\it matter} fields in the expressions for the 
(anti-)BRST charges. As a consequence, the nilpotency and absolute anticommutativity 
of the (anti-)BRST charges can be captured within the framework of ACSA to BRST formalism, too, as has been
done in our previous work [16]. In fact, we have been able to demonstrate the above {\it mathematically elegant}
properties within the framework of AVSA to  BRST formalism where the {\it full} expansions of the superfields
 have been taken into account [16]. In the proof of the absolute anticommutativity 
property, we have been forced to invoke the CF-condition [17] to recast the expressions for the  nilpotent (anti-)BRST charges
in their appropriate forms (see, Sec. 6)  as has been {\it also} done in our earlier works [16, 18].
Thus, we adopt here the same theoretical trick  for the proof of the absolute anticommutativity
of the  nilpotent (anti-)BRST charges  for our {\it interacting} non-Abelian 1-form theory with Dirac fields..

One of the highlights of our present investigation is the theoretical material contained in
Sec. 6 where we have captured the nilpotency and absolute anticommutativity properties of the 
(anti-)BRST charges within the framework of ACSA to BRST formalism. We have been able to express 
these expressions in the {\it ordinary} space where the explicit (anti-)BRST symmetry transformations for the {\it ordinary} fields 
and their off-shell {\it nilpotency} properties have been taken into account in a judicious manner.
It is pertinent, at this stage, to pinpoint the fact that our knowledge of the {\it ordinary} 
and {\it superspace} formulations have  helped each-other in a beautiful and complementary fashion in our 
theoretical discussions. Sometimes  our knowledge, in the ordinary space,
has helped us in our theoretical discussions in the context of superspace formulation and, at other times, our understanding 
of the {\it superspace} formulation has come in handy for our theoretical discussions  in the   ordinary  space.
Thus, the contents of Sec. 6  (which are one of the highlights of our present endeavor) are  
the outcome of our understandings of the BRST formalism in the {\it ordinary} space and {\it superspace}
and their inter-connections\footnote {We have established that the nilpotent and absolutely anticommuting (anti-)BRST 
  transformations (and their corresponding conserved charges) are deeply connected with the translational 
generators $(\partial_{\bar\theta}, \partial_\theta)$ along the Grassmannian directions of the ($D$, 1)-dimensional 
(anti-)chiral super-submanifolds (of the {\it general} ($D$, 2)-dimensional supermanifold on which our $D$-dimensional ordinary 
 non-Abelian 1-form {\it interacting} gauge theory is generalized).}.

Against the backdrop of the above discussions, we would like to lay emphasis on the fact that we have {\it also} applied the (anti-)chiral supervariable approach  (ACSA) to the  ${\cal N} = 2$ SUSY quantum mechanical models of various kinds [19-22] and derived the ${\cal N} = 2$ SUSY symmetry transformations and corresponding conserved and nilpotent super charges. In this derivation, the role of SUSY invariant quantities has been very decisive because we have demanded that {\it such} quantities should {\it not} depend on the Grassmannian variables $(\theta, \bar{\theta})$. We have been able to prove the nilpotency of these ${\cal N} = 2$ super charges. However, it has been found that the ACSA (applied to the ${\cal N} = 2$ SUSY QM models) does not lead to the derivation of the absolute anticommutativity for the ${\cal N} = 2$  conserved and nilpotent super charges. This observation, for obvious reasons, is consistent with the 
basic tenets of ${\cal N} = 2$ SUSY QM systems.
In our Appendix B, we demonstrate {\it this} fact in an explicit fashion so that {\it novelty} of our observation of the absolute anticommutativity property  for the (anti-)BRST charges could be corroborated within the framework 
of ACSA. Thus, it is crystal clear that ACSA does {\it not} lead to the absolute anticommutativity of the 
conserved and nilpotent charges {\it everywhere}.

The following key factors have propelled our curiosity to pursue our present investigation.
First, we have captured the nilpotency and absolute anticommutativity of the fermionic (anti-)BRST
charges for the interacting {\it Abelian} 1-form gauge theories with Dirac and complex scalar 
fields [18] as well as 4$D$ Abelian 2-form gauge theory  [15] within the framework of  ACSA to BRST
formalism. Thus, it is very important for us to prove the {\it same} in the context of an {\it interacting} 
non-Abelian 1-form gauge theory with {\it Dirac fields} so that our ideas, connected with the  ACSA to BRST 
formalism, could be firmly established. Second, the ideas of ACSA to BRST formalism are simple 
and straightforward and they lend support to the augmented version of superfield approach (AVSA) 
to BRST formalism which is based on the more {\it formal} and {\it precise} mathematical foundations (see, e.g. [4, 5, 9-12, 16]).
Our present work, once again, establishes the validity of {\it this} observation where there is 
a complete agreement between {\it our} results (with ACSA) and that of the AVSA to BRST formalism.
Third, to establish the {\it novelty} of our observation of the absolute anticommutativity property  in the context of the (anti-)BRST charges, 
we have concentrated on the application of ACSA to an ${\cal N} = 2$ SUSY QM model of harmonic oscillator and shown that the ${\cal N} = 2$ super charges do {\it not} absolutely anticommute (cf. Appendix B).
Finally, our present endeavor is  also our modest step forward  towards our central objective
 of applying the theoretical techniques of ACSA to BRST formalism in the context of higher $p$-form ($p$ = 2, 3...)
 gauge theories as well as other physically  interesting ${\cal N} = 2$ SUSY QM models (which are popular in literature).

The theoretical materials of our present investigation are organized as follows. 
First of all, we discuss in Sec. 2, the bare essentials of the nilpotent and absolutely anticommuting (anti-)BRST symmetries 
within the framework of Lagrangian formulation. The subject matter of Sec. 3 concerns itself with the derivation of 
BRST symmetries  of our theory by exploiting the {\it anti-chiral} superfields and their 
super expansions. Our Sec. 4 is devoted to the discussion of anti-BRST symmetries
which are derived by using the anti-BRST invariant restrictions on the {\it chiral} superfields.
Sec. 5 of our paper contains the discussion about the invariance of the  Lagrangian densities  within the framework
of (anti-)chiral superfield formalism. In Sec. 6, we deal with the discussion of nilpotency 
and absolute anticommutativity properties of the conserved (anti-)BRST charges within
the framework of ACSA to BRST formalism. Finally, we summarize 
our key results in Sec. 7 and point out a few possible future theoretical directions for further investigation(s).

In our Appendix A, we concisely discuss about the {\it novelty} of our key  observation of the absolute anticommutativity
property (associated with the (anti-)BRST charges) within the framework of ACSA to BRST formalism. Our Appendix B is devoted to the 
application of ACSA to the ${\cal N } = 2$ SUSY QM model of a 1$D$ harmonic oscillator where we demonstrate that the ${\cal N } = 2$ 
nilpotent and conserved super charges do {\it not}
absolutely anticommute.\\

\noindent
{\it Convention and Notations}: We adopt the convention of taking the metric tensor $\eta_{\mu\nu}$ for the
background $D$-dimensional flat Minkowskian spacetime as: $\eta_{\mu\nu} =$ diag $(+1, -1, -1...)$ so that
$\partial_\mu A^\mu  = \partial_0 A_0 - \partial_i A_i$ where the Greek indices $\mu, \nu, \lambda... = 0, 1, 2...D-1$
represent the time and space directions and Latin indices $i, j, k... = 1, 2, 3...D-1$ correspond to the {\it space}
directions only. We choose the convention of dot and cross products in the Lie algebraic space as: 
$P\cdot Q = P^a Q^a$ and $(P\times Q)^a = f^{abc}P^b Q^c$ between a set of two non-null vectors ($P^a, Q^a$)
where $a, b, c... = 1, 2, 3...N^2 - 1$ and $f^{abc}$ are the 
totally antisymmetric structure constants
for the $SU(N)$ Lie algebra. We have also adopted the convention of left-derivative in {\it all} our
relevant computations with respect to the fermionic fields ($\psi,\,\bar\psi, C, \bar C$) which obey:
$\psi\,\bar\psi+\bar\psi\,\psi = 0, \psi^2 = 0, \bar\psi^2 = 0, (C)^2 = 0, (\bar C)^2 = 0, C^{a}\;\bar C^{b} + \bar C^{b}\;C^{a} = 0,
C^{a}\;C^{b} +  C^{b}\;C^{a}  = 0, \bar C^{a}\;\bar C^{b} + \bar C^{b}\;\bar C^{a} = 0,$ 
 $C\,\psi + \psi\, C = 0,$ etc. We denote the (anti-)BRST transformations and corresponding conserved charges  by the notations $s_{(a)b}$ 
and  $Q_{(ab)}$ in the whole 
body of our text. In the context of the  ${\cal N } = 2$ SUSY QM model,  we have adopted  the notations $s_1$ and $s_2$ for the SUSY transformations and corresponding conserved charges have been denoted  by $Q$ and $\bar Q$.

\section{Preliminaries: Lagrangian Formulation }

We begin with the (anti-)BRST invariant coupled (but equivalent) Lagrangian densities (see, e.g. [23] for details) for the $D$-dimensional 
non-Abelian 1-form gauge theory where there is a coupling between the gauge field ($A_\mu$) and Dirac fields ($\bar\psi,  \psi$),   
in the Cruci-Ferrari gauge [24, 25], as 
\begin{eqnarray}
{\cal L}_B  & = & -\,\frac{1}{4}F^{\mu\nu}\cdot F_{\mu\nu} + \bar\psi\,(i\,\gamma ^ \mu D_\mu - m)\,\psi + B\cdot (\partial_\mu A^\mu)\nonumber\\
& + & \frac{1}{2}(B\cdot B + \bar B\cdot \bar B) - i\,\partial_{\mu}\bar C\cdot D^{\mu} C,\nonumber\\
{\cal L}_{\bar B}  & = & -\,\frac{1}{4}F^{\mu\nu}\cdot F_{\mu\nu} + \bar\psi\,(i\,\gamma ^ \mu D_\mu - m)\,\psi - \bar B\cdot (\partial_\mu A^\mu)\nonumber\\
 & + & \frac{1}{2}(B\cdot B + \bar B\cdot \bar B)
 -  i\,D_{\mu}\bar C\cdot \partial^{\mu} C, 
\end{eqnarray}
where the covariant derivatives $D_\mu\psi = \partial_\mu\psi  + i\,(A_\mu\cdot T)\,\psi$ and $D_\mu C = \partial_\mu C  + i\,(A_\mu\times C)$
are in the fundamental and adjoint representations of the $SU(N)$ Lie algebra, respectively. This algebra
 is generated by the operators $(T^a)$ that satisfy: 
$[T^a, T^b] = f^{abc}\;T^c$ where $f^{abc}$ are the structure constants that can be chosen to be totally antisymmetric in indices
$a, b, c = 1, 2...N^2 - 1$ for the semi-simple Lie group $SU(N)$ (see, e.g. [26] for details).

In the above, the Nakanishi-Lautrup type auxiliary fields $B(x)$ and $\bar B(x)$ satisfy the  
Curci-Ferrari (CF)-condition $ B + \bar B + (C\times\bar C) = 0$ [17] which emerges from the equivalency requirement 
of the coupled ({\it but} equivalent) Lagrangian densities ${\cal L}_B$ and ${\cal L}_{\bar B}$ that mathematically  implies the following
\begin{eqnarray}
&& B\cdot (\partial_\mu A^\mu) - i\,\partial_\mu\bar C\cdot D^\mu C \equiv  - \bar B\cdot (\partial_\mu A^\mu) - i D_\mu\bar C\cdot \partial^\mu C,
\end{eqnarray}
modulo a total spacetime derivative. It turns out,  the following infinitesimal, continuous, off-shell nilpotent $(s_{(a)b}^2 = 0)$
and absolutely anticommuting $(s_b s_{ab} + s_{ab} s_b = 0)$ (anti-)BRST symmetry transformations $(s_{(a)b})$
\begin{eqnarray*}
&&s_{ab}\; A_\mu= D_\mu\bar C,\qquad  s_{ab}\; \bar C= -\frac{i}{2}\,(\bar C\times\bar C),
\qquad s_{ab}\;C   = i{\bar B},\qquad  s_{ab}\;\bar B    = 0,\nonumber\\
&&s_{ab}\; F_{\mu\nu}      = i \,(F_{\mu\nu}\times\bar C),\qquad s_{ab} (\partial_\mu A^\mu) = \partial_\mu D^\mu\bar C,
\qquad s_{ab}\;\psi = -\;i\;\bar C\;\psi,\nonumber\\
&& s_{ab}\;\bar\psi = -\;i\;\bar\psi\;\bar C,
\qquad s_{ab}\; B = i\,(B \times \bar C), \nonumber\\
&&s_b\; A_\mu = D_\mu C,\qquad s_b \;C =  - \frac{i}{2}\; (C\times C),\qquad  s_b\;\bar C \;= i\,B ,\;
 \qquad s_b\; B = 0, \nonumber\\
 \end{eqnarray*}
\begin{eqnarray}
&& s_b\;\bar B = i\,(\bar B\times C),\qquad s_b\; (\partial_\mu A^\mu) = \partial_\mu D^\mu C,
\qquad s_b\;\psi = -\;i\; C\;\psi,\nonumber\\
&& s_b\;\bar\psi = -\;i\;\bar\psi\; C,\qquad s_b \;F_{\mu\nu} = i\,(F_{\mu\nu}\times C),
\end{eqnarray}
leave the action integrals ($ S_1 = \int \,d^{D} x\, {\cal L}_B $ and $S_2 = \int \,d^{D} x\, {\cal L}_{\bar B}$) invariant. In fact,
we also note here that the Lagrangian densities transform to the total spacetime derivatives (plus extra terms) under $s_{(a)b}$ as given below:
\begin{eqnarray}
s_b{\cal L}_B  & = & \partial_\mu(B \cdot D^\mu C),  \quad s_{ab}{\cal L}_{\bar B}= - \;\partial_\mu{(\bar B \cdot D^\mu \bar C)},\nonumber\\
s_b{\cal L }_{\bar B}\; & = & \partial_\mu\,[ {\{ B + ( C \times \bar C )\}}\cdot
\partial^\mu C \,]-{\{B + \bar B + ( C\times\bar C )\}}\cdot D_\mu\partial^\mu C,\nonumber\\
s_{ab}{\cal L}_B & = &-\;\partial_\mu\,[{\{\bar B + ( C\times\bar C)\} \cdot \partial^\mu \bar C}\,] 
 +  \;\{(B+\bar B + ( C \times {\bar C})\} \cdot D_\mu \partial^\mu \bar C.
\end{eqnarray}
We point  out that the above extra pieces, besides the total spacetime derivative terms, are deeply  
connected with the CF-condition: $B + \bar B + ( C\times\bar C) = 0$.

At this  juncture, a few comments are in order. First of all, we note 
that {\it both} the Lagrangian densities respect {\it both} (BRST and anti-BRST) symmetry  
transformations $(s_{(a)b})$ if the whole {\it interacting} non-Abelian theory is confined to be defined on
a hypersurface in the $D$-dimensional Minkowski space where the CF-condition $[B + \bar B + ( C\times\bar C) = 0]$ 
is satisfied. In other words, we note that we have the following
\begin{eqnarray}
&&s_b\;{\cal L}_{\bar B} = -\partial_\mu[{\bar B}\cdot\partial^{\mu}C],\qquad\qquad 
s_{ab}\,{\cal L}_{B} = \partial_\mu[B \cdot\partial^{\mu}\bar C ],
\end{eqnarray}
on the constrained hypersurface where $B + \bar B + ( C\times\bar C) = 0$. Second,
as is evident from Eq. (2), both the Lagrangian densities are equivalent {\it only}
on the hypersurface, defined on the $D$-dimensional Minkowiskian spacetime manifold, by the 
CF-condition. Finally, we can explicitly check that the (anti-)BRST symmetry transformations
$(s_{(a)b})$ are absolutely anticommuting {\it only} on the {\it above} hypersurface because it can 
be proven that the anticommutators
\begin{eqnarray}
&& {\{s_b, s_{ab}}\}\;A_\mu  = 0,\quad\; {\{s_b, s_{ab}}\}\; F_{\mu\nu} = 0,\quad\; {\{s_b, s_{ab}}\}\;\psi = 0,\quad\; {\{s_b, s_{ab}}\}\;\bar\psi = 0,
\end{eqnarray}
are satisfied only when the CF-condition is taken into account.

According to the Noether theorem, the invariance of the action integrals 
$(S_1 = \int d^D x \,{\cal L}_B, S_2 = \int d^D x\, {\cal L}_{\bar B}),$ 
under the BRST and anti-BRST transformations (i.e. $s_{b}\;{\cal L}_ B  = \partial_\mu [B\cdot D^\mu C], s_{ab}\;{\cal L}_{\bar B}  
= -\,\partial_\mu [\bar B\cdot D^\mu \bar C]$), leads to the following Noether conserved currents:
\begin{eqnarray}
&&J^{\mu}_{(ab)} = -\,\bar B\cdot D^\mu \bar C - F^{\mu\nu}\cdot D_\nu \bar C - \bar\psi\;\gamma^{\mu}\;\bar C\;\psi - 
\frac {1}{2}\;(\bar C\times \bar C)\cdot {\partial^\mu C},\nonumber\\
&&J^{\mu}_{(b)} = B\cdot D^\mu C - F^{\mu\nu}\cdot D_\nu C - \bar\psi\;\gamma^{\mu}\;C\;\psi + \frac {1}{2}\;\partial^\mu \bar C\cdot (C\times C).
\end{eqnarray} 
The conservation laws ($\partial_{\mu}J_{(a)b}^{\mu} = 0$) can be proven by using the following 
Euler-Lagrange (EL) equations  of motion (EOM) which emerge  from the Lagrangian density ${\cal L}_B$, namely; 
\begin{eqnarray}
&&B = -\;(\partial_\mu A^\mu),\;\;\qquad \partial_\mu (D^\mu C) = 0,\;\;\qquad D_\mu (\partial^\mu\bar C) =0,\nonumber\\
&&(i\,\gamma^\mu\partial_\mu  - m)\psi  = \gamma^\mu\; A_\mu\; \psi,\qquad
i\;(\partial_\mu \bar\psi)\gamma^\mu + m\bar\psi = -\;\bar\psi\;\gamma^\mu\; A_\mu,\nonumber\\
&&D_\mu F^{\mu\nu} - \partial^\nu B = \bar\psi\;\gamma^\nu\;\psi  + (\partial^\nu \bar C\times C). 
\end{eqnarray}
In {\it an} exactly similar fashion, one could compute the EL-EOM from the Lagrangian density ${\cal L}_{\bar B}$
which turn out to be similar to (8) except the following: 
\begin{eqnarray}
&&D_\mu F^{\mu\nu} + \partial^\nu \bar B \; = \;  \bar\psi\;\gamma^\nu\;\psi  - (\bar C\times \partial^\nu C),
\nonumber\\
&&\partial_\mu (D^\mu \bar C) = 0,\qquad D_\mu (\partial^\mu C) =0 , \qquad \bar B = \partial_\mu A^\mu.
\end{eqnarray}
The conserved (anti-)BRST charges ($Q_{(a)b} = \int d^{D-1} x\;J{^0}_{(a)b}$)  can be derived from the above 
 conserved currents (7) as:
\begin{eqnarray}
&&Q_{ab}= \int d^{D-1} x \;\Big[-\,\bar B\cdot D_0 \bar C - \frac{1}{2}\;(\bar C\times \bar C)\cdot{C} 
-\bar\psi\;\gamma^0\; \bar C\;\psi - F^{0i}\cdot D_i \bar C\Big],\nonumber\\
&&Q_b  = \int d^{D-1} x \;\Big[B\cdot D_0 C + \frac{1}{2}\;\dot{\bar C}\cdot (C\times C)-\bar\psi\;\gamma^0\; C\;\psi
 - \,F^{0i}\cdot D_i C\Big].
\end{eqnarray}
It can be explicitly checked that the terms ($- \,F^{0i}\cdot D_i C$) and ($- \,F^{0i}\cdot D_i \bar C$) can be written, in terms 
of total {\it space} derivatives,  as:
\begin{eqnarray}
- \,F^{0i}\cdot D_i C = -\;\partial_i\; [F^{0i}\cdot C] + (D_i F^{0i})\cdot C,\nonumber\\
- \,F^{0i}\cdot D_i \bar C = -\;\partial_i \;[F^{0i}\cdot \bar C] + (D_i F^{0i})\cdot \bar C.
\end{eqnarray}
Applying the Gauss divergence theorem and using the EL-EOM w.r.t. gauge field, it is straightforward  to 
check that the above charges can be re-expressed in the following concise forms:
\begin{eqnarray}
&&Q_{ab}  = \int d^{D-1} x\;\Big[\dot{\bar B}\cdot \bar C -\,\bar B\cdot D_0 \bar C + \frac {1}{2}\;(\bar C\times \bar C)\cdot \dot C \Big],\nonumber\\
&&Q_b  = \int d^{D-1} x\;\Big[B\cdot D_0 C - \dot B\cdot C - \frac {1}{2}\;\dot{\bar C}\cdot (C\times C)\Big].
\end{eqnarray}
To be more precise, we have used the following EL-EOM
\begin{eqnarray}
&&D_i F^{i0} -\dot B = \bar\psi\; \gamma^0\;\psi + (\dot {\bar C}\times C),\nonumber\\
&& D_i F^{i0} -\dot {\bar B} = \bar\psi\; \gamma^0\;\psi - ( {\bar C}\times\dot C),
\end{eqnarray}
which have emerged out from Eqs. (8) and (9).

It is interesting to point out that the matter fields {\it disappear} from 
the {\it final} expressions for the above (anti-)BRST charges and these expressions 
appear as if there were {\it no} matter fields in the theory.
This has happened because of the fact that we have used the EOM (13) and the theoretical technique elaborated in (11).
 We shall concentrate on these concise expressions (i.e. Eq. (12)) of the conserved charges 
for our further discussions within the framework of ACSA to BRST 
formalism and capture their nilpotency and absolute anticommutativity 
properties.

\noindent
\section{Off-Shell Nilpotent BRST Symmetries: Anti-Chiral Superfields  and Their Super Expansions}

We exploit, in this section, the strength of  BRST (i.e. quantum gauge) invariant restrictions on the anti-chiral 
superfields to derive the off-shell nilpotent BRST symmetry transformations. Towards this goal in mind, 
first of all, we generalize the ordinary $D$-dimensional fields 
 ($A_\mu, C, \bar C, B, \bar B, \psi, \bar\psi$) of the Lagrangian density (1)
onto a $(D, 1)$-dimensional anti-chiral super-submanifold (parameterized by the  superspace coordinates $x^\mu$ and $\bar\theta$) as 
\begin{eqnarray}
&&A_{\mu} (x) \,\longrightarrow~  B_\mu (x, \bar\theta)  =  A_{\mu}(x) + \bar\theta\,  R_\mu (x),\quad
C (x)\;\; \longrightarrow~ F(x, \bar\theta)  =  C(x) + i\,\bar\theta\,  B_{1}(x),\nonumber\\
&&\bar C (x) \;\;\longrightarrow~ \bar F(x, \bar\theta)  =  \bar C_(x) + i\,\bar\theta\,  B_{2} (x),\quad
\;\;B (x)\;\;  \longrightarrow~ \tilde B (x, \bar\theta) = B(x) + i\,\bar\theta\,  f{_1}(x),\nonumber\\
&&\bar B (x)\;\;  \longrightarrow~ \tilde{\bar B} (x, \bar\theta) = \bar B(x) + i\,\bar\theta\,  f{_2}(x),
\quad\;\;\psi (x)\;\;\,  \longrightarrow ~\Psi  (x, \bar\theta)  =  \psi (x) + i\,\bar\theta\;b_1 (x),\nonumber\\
&&\bar \psi (x) \;\;\,\longrightarrow ~\bar \Psi(x, \bar\theta)  =  \bar \psi (x) + i\,\bar\theta\,  b_{2} (x),
\end{eqnarray}
where the secondary fields ($B_1, B_2, b_1, b_2$), on the r.h.s.,  are bosonic in nature and the set of secondary fields 
($R_\mu, f_1, f_2$) is fermionic. We further note that all the fields are defined as: $A_\mu = A_\mu\cdot T$, $C = C \cdot T$, 
$b_1 = b_1 \cdot T$, $R_\mu = R_\mu \cdot T,$ etc. We have to derive  the explicit form of the above secondary fields in terms of the {\it basic} and 
{\it auxiliary} fields of our starting Lagrangian densities (1) for our non-Abelian 1-form {\it interacting} gauge theory
in $D$-dimensions of spacetime.

Towards the above objective in mind, we list here the useful and interesting  BRST invariant 
quantities for the Lagrangian density ${\cal L}_B $ (of (1)) as:
\begin{eqnarray}
&& s_b \;(D_\mu C) = 0, \qquad s_b \;B = 0, \qquad s_b \;(C\times C) = 0,\qquad s_b\; (C\;\psi) = 0,\nonumber\\
&& s_b\; (\bar\psi \; C) = 0,\qquad s_b \;\big[A^\mu\cdot\partial_\mu B + i\;\partial_\mu\bar C\cdot D^\mu C\big] = 0.
\end{eqnarray}
According to the basic tenets of ACSA to BRST formalism, the above set of quantities\footnote{The analogues of 
the GIRs (at the {\it classical level}) are the (anti-)BRST invariant quantities (at the {\it quantum} level). Hence, they are {\it physical} 
and they are required to be independent of the ``soul" coordinates when they are recast in the language of superfields 
within the framework of ACSA to BRST formalism. We would like to mention here that the BRST
invariant quantities in (15) have been found by the method of trial and error as there is no definite rule/principle
to obtain them.} should be
independent of the ``soul" coordinate  $\bar\theta$ when these physically important quantities are generalized 
onto a ($D$, 1)-dimensional anti-chiral super-submanifold (of the {\it general} ($D$, 2)-dimensional 
supermanifold). For instance, we note that, the following are true
\footnote{We shall denote the anti-chiral superfields with superscript $(b)$ whose super expansions would lead to the derivation
of BRST symmetry transformations (for the corresponding  {\it ordinary} fields) as the coefficient of $\bar\theta$ in the 
anti-chiral super expansions of the superfields.}, namely;
\begin{eqnarray}
&& s_b B = 0~~~~~~\,~~\Longrightarrow  \tilde B (x, \bar\theta) = B(x) \Longrightarrow  f_1 (x) = 0\nonumber\\
&&~~~~~~~~~~~~~~~~~~~\Longrightarrow \tilde B^{(b)}(x,\bar\theta) = B(x) + \bar\theta\; (0),\nonumber\\
&& s_b (C\times C) = 0 \Longrightarrow F(x,\bar\theta)\times F(x, \bar\theta) = C(x)\times C(x)\nonumber\\
&& ~~~~~~~~~~~~~~~~~~~\Longrightarrow B_1\times C = 0.
\end{eqnarray}
The latter condition $B_1\times C = 0 $ implies that {\it one} of the non-trivial solutions of this restriction  
is $B_1\propto (C\times C)$ because we know that: $(C\times C)\times C = 0$. Let us choose $B_1 = \kappa\; (C\times C)$
where $\kappa $ is some numerical constant. With the above choice, we have the reduced/modified form 
of the superfield $F(x, \bar\theta)$ as:
\begin{eqnarray}
&& F (x, \bar\theta)\longrightarrow F^{(m)}(x,\bar\theta) = C(x) + i\;\bar\theta\; \kappa\; (C\times C).
\end{eqnarray}
Here the superscript $(m)$ denotes the modified  form of the superfield.
Now we focus on $s_b (D_\mu C) = 0$ which implies the following equality (with the input from (17)):
\begin{eqnarray}
&&\partial_\mu F^{(m)} (x, \bar\theta) + i\, B_\mu (x, \bar\theta)\times F^{(m)}(x,\bar\theta) = \partial_\mu C(x) + i\, A_\mu (x)\times C(x).
\end{eqnarray}
The substitution of the super expansions from (14) and (17), in the above, leads to the following important relationship:
\begin{eqnarray}
&& R_\mu  = -\; 2 \;\kappa \; D_\mu C (x).
\end{eqnarray}
As a consequence, we have the modified  form of the   superfield $B_\mu (x, \bar\theta)$ as:
\begin{eqnarray}
&& B_\mu (x,\bar\theta)\longrightarrow B^{(m)}_\mu (x,\bar\theta) = A_\mu (x)  -\; 2 \;\kappa \;\bar\theta\; D_\mu C (x).
\end{eqnarray}
We exploit now the BRST invariant quantity $s_b (B\times\bar C) = 0$. This invariance leads to the following equality
(with input from (16)), namely;
\begin{eqnarray}
&&\tilde B^{(b)}(x,\bar\theta)\times \bar F(x,\bar\theta) = B(x)\times\bar C(x),
\end{eqnarray}
and using (15) and (14), we obtain $B_2(x)\propto B(x)$. For the sake of brevity\footnote{We have chosen $B_2 (x) = B(x)$ to be
consistent with the BRST symmetry transformations (3). However, we have the freedom to choose 
$B_2 (x) = \alpha \,B(x)$ where $\alpha $ is a constant numerical factor.}, however, we choose $B_2(x) = B(x)$
so that we obtain 
\begin{eqnarray}
&& \bar F^{(b)}(x, \bar\theta) = \bar C(x) + i\;\bar\theta \,B(x) = \bar C(x) + \bar\theta\; (s_b \bar C(x)),
\end{eqnarray}
where the superscript $(b)$ denotes that the above superfield has been obtained after the BRST
invariant restriction (21). We use now the following equality
\begin{eqnarray}
 &&B^{\mu(m)}(x, \bar\theta)\cdot \partial_\mu \tilde B^{(b)} (x, \bar\theta) + i\;\partial_\mu \bar F^{(b)} (x, \bar\theta)\cdot\partial^\mu F^{(m)}(x,\bar\theta)\nonumber\\
& - &\partial_\mu \bar F^{(b)} (x, \bar\theta)\cdot \Big[B^{\mu (m)}(x,\bar\theta)\times F^{(m)}(x,\bar\theta)\Big]\nonumber\\
& = & A^\mu(x)\cdot\partial_\mu B(x) + i\, \partial_\mu\bar C(x)\cdot\partial^\mu C(x)\nonumber\\
& - & \partial_\mu\bar C(x)\cdot \Big[A^\mu (x)\times C(x)\Big],
\end{eqnarray}
which emerges from the BRST invariant quantity $s_b\;[A^\mu\cdot \partial_\mu B(x) + i\;\partial_\mu\bar C(x)\cdot D^\mu C(x)] = 0$.
Substitutions of the super expansions from (14), (17), (20) and (22) lead to:
\begin{eqnarray}
&&\kappa    =  -\;\frac {1}{2}\;\;\Longrightarrow \;\;B^{(b)}_\mu(x, \bar\theta) = A_\mu (x) + \bar\theta \;(D_\mu C(x))\equiv A_\mu (x) 
+ \bar\theta \;(s_b A_\mu (x)),\nonumber\\
&&\kappa  =  -\;\frac {1}{2}\;\;\Longrightarrow \;\; F^{(b)}(x,\bar\theta) =  C(x) + \bar\theta\;\Big[-\;\frac {i}{2}\; (C\times C)\Big]\nonumber\\
&& ~~~~~~~~~~~~~~~~~~~~~~~~~~~~~~~~~\equiv  C(x) + \bar\theta\; (s_b C(x)).
\end{eqnarray}
We note (from (22), (24) and (14)) that we have already  derived the BRST transformations: $s_b\bar C = iB,\; s_b A_\mu = D_\mu C,\; s_b C = 
-  \;(\frac {i}{2})  \;(C\times C),\;
s_b B = 0$ as the coefficients of $\bar\theta$ in the super expansions of the anti-chiral superfields with superscript $(b)$
which are derived after the applications of  BRST invariant restrictions.

We now focus on the derivation of $s_b \bar B = i\; (\bar B\times C),\; s_b \psi = - \,i\; C\;\psi,\; s_b\;\bar\psi  = -i\;\bar\psi\; C$ from the 
BRST symmetry invariances:  $s_b (\bar B\times C) = 0,\; s_b (C\;\psi) = 0$ and $s_b (\bar\psi\; C) = 0$.
These invariances lead to the following equalities  and their consequences (i.e. the derivation of secondary fields), namely; 
\begin{eqnarray}
&&\tilde {\bar B}(x,\bar\theta)\times F^{(b)}(x,\bar\theta) = \bar B(x)\times C(x)\Longrightarrow  f_2 (x)  =  (\bar B\times C),\nonumber\\
&&F^{(b)}(x,\bar\theta) \;\Psi (x,\bar\theta) = C(x)\; \psi (x) \qquad \Longrightarrow b_1(x) = -\;C(x)\;\psi (x),\nonumber\\
&&\bar \Psi (x,\bar\theta)\; F^{(b)}(x,\bar\theta) = \bar \psi (x)\; C(x)\quad\quad\Longrightarrow  b_2(x) = -\;\bar\psi(x)\; C(x).
\end{eqnarray}
We would like to mention  here that we have used: $C\;C = \frac {1}{2}\;{\{C, C}\} \equiv \frac {1}{2}\; (C\times C)$
in the derivation of the  expressions for $b_1 (x)$ and $b_2 (x)$. 
The substitutions of these values into the super expansions (14) lead to the following expansions
\begin{eqnarray}
&&{\bar B}^{(b)}(x,\bar\theta) = \bar B(x) + \bar\theta\; [i\;(\bar B\times C)] =  \bar B(x) + \bar\theta\; (s_b \bar B(x)),\nonumber\\
&&\Psi^{(b)}(x,\bar\theta) = \psi (x) + \bar\theta\; (-\;i \; C\;\psi) \equiv \psi (x) + \bar\theta\; (s_b \psi(x)),\nonumber\\
&&{\bar\Psi}^{(b)}(x,\bar\theta) = \bar \psi (x) + \bar\theta\; (-\;i\;\bar\psi\; C)\equiv \bar\psi (x) + \bar\theta\; (s_b\bar\psi(x)),
\end{eqnarray}
where superscript $(b)$ denotes the anti-chiral superfields that have been obtained after the applications of BRST invariant restrictions.

We end this section with the remark that we have obtained {\it all} the BRST symmetry transformations for {\it all}
the fields of our present non-Abelian 1-form gauge theory (where there is an {\it interaction} between the gauge field 
and Dirac fields ($\psi, \bar\psi$)) as the coefficient of $\bar\theta$ in the super expansions of the anti-chiral superfields 
with superscript $(b)$. We lay emphasis on the fact that our results lend support to the results obtained in
[4, 5] which are obtained by the formal application of the HC (that depends crucially on the 
exterior derivative  $d = dx^\mu\partial_\mu \;(d^2 = 0)$ of the  de Rham cohomological operators of differential geometry). 
In addition, we derive the off-shell nilpotent (anti-)BRST symmetry transformations for the {\it matter} fields which has not been derived
 and discussed in [4, 5].

\noindent
\section{Off-Shell Nilpotent Anti-BRST Symmetries: Chiral Superfields   and Their Chiral Super Expansions}In this section, we derive the anti-BRST symmetry transformations for our interacting non-Abelian 1-form gauge theory by 
invoking the anti-BRST (i.e. quantum gauge) invariant restrictions on the {\it chiral} superfields. Towards this goal in mind,
we generalize the  ordinary fields $(A_\mu, C, \bar C, B, \bar B, \psi, \bar\psi)$ of our $D$-dimensional  {\it ordinary}
non-Abelian 1-form gauge  theory onto 
the {\it chiral} ($D$, 1)-dimensional super-submanifold (of the {\it general} ($D$, 2)-dimensional supermanifold) as
\begin{eqnarray}
&&A_{\mu} (x)\, \longrightarrow ~ B_\mu (x, \theta)  =  A_{\mu}(x) + \theta\, \bar R_\mu (x),\quad
C (x) \;\;\longrightarrow~ F(x, \theta)  =  C(x) + i\,\theta\;\bar  B_{1}(x),\nonumber\\
&&\bar C (x) \;\;\longrightarrow ~ \bar F(x, \theta)  =  \bar C_(x) + i\,\theta\, \bar B_{2} (x),\quad
\;\;B (x)\;\;  \longrightarrow ~ \tilde B (x, \theta) = B(x) + i\,\theta\,\bar  f{_1}(x),\nonumber\\
&&\bar B (x)\;\;  \longrightarrow ~ \tilde{\bar B} (x, \theta) = \bar B(x) + i\,\theta\,\bar  f{_2}(x),\quad
\;\;\psi (x)\;\;\,  \longrightarrow ~\Psi  (x, \theta)  =  \psi (x) + i\,\theta\;\bar b_1 (x),\nonumber\\
&&\bar \psi (x) \;\;\,\longrightarrow ~ \bar \Psi(x, \theta)  =  \bar \psi (x) + i\,\theta\, \bar b_{2} (x),
\end{eqnarray} 
where we point out explicitly that the ($D$, 1)-dimensional {\it chiral} super-submanifold is parameterized by the bosonic coordinates $x^\mu \; (\mu = 0, 1
.....D-1)$ and fermionic $(\theta^2 = 0)$ Grassmannian variable $\theta$. To obtain the secondary fields $(\bar R_\mu (x), \bar  B_{1}(x),
\bar  B_{2}(x), \bar  f_{1}(x), \bar  f_{2}(x), \bar  b_{1}(x), \bar  b_{2}(x))$ in terms of the basic and auxiliary fields of 
the Lagrangian densities (1), we have found out (by the method of trial and error) the following useful and interesting  anti-BRST 
invariant quantities
\begin{eqnarray}
&& s_{ab} (\bar C\;\psi) = 0,\; \; \;\;
s_{ab} \;(D_\mu\bar C) = 0, \;\; \;\,s_{ab} \;(\bar B) = 0, \;\; \;\, s_{ab} \;(\bar C\times \bar C) = 0,\nonumber\\
&& s_{ab}\; (B\times \bar C) = 0,\; s_{ab} \;[A^\mu\cdot\partial_\mu\bar B + i\, D^\mu \bar C\cdot \partial_\mu C ] = 0 ,\, s_{ab} (\bar\psi
\,\bar C) = 0,
\end{eqnarray}
where the fields (in the brackets) are present in the starting Lagrangian density (${\cal L}_{\bar B}$) that respects {\it perfect}
anti-BRST symmetry  (i.e. $s_{ab}\; {\cal L}_{\bar B} = -\;\partial_\mu [ \bar B\cdot D^\mu\bar C]$) in the sense that the corresponding 
action integral ($S = \int d^{D}x \;{\cal L}_{\bar B}$) remains {\it invariant} for the physically well-defined fields which
vanish off at $x = \pm \infty$.

We do not elaborate  here  on {\it all} the step-by-step computations (as we have done in the 
previous section). The algebraic computations are exactly on the similar lines as in the previous section. Thus, we collect here 
{\it all} the key results that emerge  out by demanding the validity of the basic tenets of ACSA to BRST formalism
where the anti-BRST (i.e. quantum gauge) invariant quantities  are required to remain independent of the ``soul" coordinate $\theta,$
 namely;
\begin{eqnarray}
&&s_{ab} \bar B(x)  = 0~~~~ \Longrightarrow \bar f_2 (x) = 0\Longrightarrow \tilde {\bar B}^{(ab)}(x,\theta)
 =  \bar B(x) + \theta \;(0)\nonumber\\
 &&~~~~~~~~~~~~~~~~~~~~\equiv   \bar B(x) + \theta \;(s_{ab} \bar B (x)),\nonumber\\
&&s_{ab} (\bar C\times \bar C) = 0\Longrightarrow \bar B_2 = \kappa \; (\bar C\times \bar C)\nonumber\\ 
&&~~~~~~~~~~~~~~~~~~~~\Longrightarrow \bar F^{(m)}(x,\theta) 
\equiv \bar C(x) + i\;\theta\;\kappa \;(\bar C\times \bar C),\nonumber\\
&&s_{ab} (\bar B \times C) = 0 \Longrightarrow
\bar B_1 (x)  = \bar B (x)\nonumber\\
&&~~~~~~~~~~~~~~~~~~~~\Longrightarrow F^{(ab)}(x, \theta) 
\equiv  C(x) + \theta\;(i\;\bar B(x)),\nonumber\\
&&s_{ab} (B\times\bar C) = 0\Longrightarrow \bar f_1(x) = B \times \bar C\nonumber\\
&&~~~~~~~~~~~~~~~~~~~~\Longrightarrow \tilde B^{(ab)} (x,\theta) = B(x) + (i\; B\times\bar C),\nonumber\\
&&s_{ab}(D_\mu \bar C) = 0\;\;\,\Longrightarrow \bar R_\mu  = -\;2\;\kappa\; (D_\mu\bar C)\nonumber\\
&&~~~~~~~~~~~~~~~~~~~~\Longrightarrow B^{ (m)}_\mu (x,\theta) = A_\mu (x) +\theta\; (-\;
2\;\kappa \; D_\mu\bar C), 
\end{eqnarray} 
where the superscripts $(m)$ and $(ab)$ denote the  modified  form of the superfields  and the  superfields that have been 
obtained after the anti-BRST invariant restrictions  which lead to the
derivation of anti-BRST symmetry transformations as the coefficients of $\theta$, respectively. We now take up the anti-BRST invariant quantity: $s_{ab}\;[ A^\mu\cdot\partial_\mu\bar B + i\; D_\mu\bar C\cdot \partial^\mu C] = 0$ which leads to the following restriction
on the {\it chiral} superfields, namely;
\begin{eqnarray}
&& B^{\mu(m)}(x, \theta)\cdot \partial_\mu \tilde{\bar B}^{(ab)} (x, \theta) + i\;\partial_\mu \bar F^{(m)} (x, \theta)\cdot\partial^\mu F^{(ab)}(x,\theta)\nonumber\\
&& -\; (B^{\mu (m)}(x,\theta)\times {\bar F}^{(m)}(x,\theta))\cdot \partial_\mu  F^{(ab)} (x, \theta)\nonumber\\
&& \equiv  A^\mu(x)\cdot\partial_\mu \bar B(x) + i\; D_\mu\bar C(x)\cdot \partial^\mu C(x),
\end{eqnarray}
where the explicit super expansions from Eq. (29) have to be used. Once  it is done, we obtain $\kappa  = -\;\frac {1}{2}$ (cf. Sec. 3)
which leads to the following: 
\begin{eqnarray}
 {\bar F}^{(m)}(x, \theta)  \longrightarrow \bar F^{(ab)}(x,\theta) & = & \bar C(x) + \theta \;(-\;\frac {i}{2}\;\bar C\times\bar C)\nonumber\\
\qquad\qquad\qquad\qquad\qquad\quad\;\;\;&\equiv &
\bar C(x) + \theta \; (s_{ab}\;\bar C(x)),\nonumber\\
 B_{\mu}^{(m)}(x,\theta)\longrightarrow B_{\mu}^{(ab)}(x,\theta) &= &A_\mu (x) + \theta\; (D_\mu\bar C(x))\nonumber\\
\qquad\qquad\qquad\qquad\qquad\quad\;\;\;&\equiv &A_\mu (x) + \theta \;(s_{ab}\; A_\mu (x)).
\end{eqnarray}
Thus, we note from Eqs. (29) and (31), that we have already derived the anti-BRST symmetry transformations for 
$\bar B(x), C(x), B(x), \bar C(x)$ and $A_\mu (x)$ as the coefficients of the Grassmannian variable $\theta$
in the {\it chiral} super expansions of the chiral superfields with the superscript $(ab)$ which are obtained
after the anti-BRST invariant restrictions which are imposed on the superfields.

We are in the position now to derive the anti-BRST symmetry transformations that are associated
with matter fields ($\psi, \bar\psi$). Towards this goal in mind, we focus on the anti-BRST invariant quantities 
$s_{ab}(\bar C\;\psi) = 0$, $s_{ab}(\bar\psi\;\bar C) = 0$ and demand the following restrictions on the composite
chiral superfields, namely;
\begin{eqnarray}
{\bar F}^{(ab)}(x, \theta) \;\Psi (x, \theta) = \bar C(x)\;\psi (x),\qquad \bar\Psi (x, \theta){\bar F}^{(ab)}(x, \theta) = \bar\psi (x)\;\bar C(x),
\end{eqnarray}
which lead to the expressions for the secondary fields $\bar b_1 (x)  = -\;\bar C(x)\;\psi (x)$ and
$\bar b_2 (x)  = -\;\bar\psi (x)\; \bar C(x)$, respectively. It is pertinent to point out
that we have used here the theoretical trick:  $\bar C\;\bar C = \frac {1}{2}\; {\{\bar C, \bar C}\} = \frac {1}{2} (\bar C\times\bar C)$
 in the determination of $b_1(x)$ and $b_2(x)$.
 These values, ultimately, lead to the following super expansions:
\begin{eqnarray}
&&\Psi^{(ab)}(x, \theta) = \psi (x) + \theta\; (-\;i \; \bar C\;\psi) \equiv \psi (x) + \theta\; (s_{ab} \psi),\nonumber\\
&&{\bar\Psi}^{(ab)}(x, \theta) = \bar \psi (x) + \bar\theta\; (-\;i\;\bar\psi\; \bar C)\equiv \bar\psi (x) + \bar\theta\; (s_{ab}\bar\psi).  
\end{eqnarray}
Thus, we have computed {\it all} the secondary fields of the super expansions (27) in terms 
of the  {\it basic} and {\it auxiliary} fields of the Lagrangian densities (1) and derived the off-shell nilpotent anti-BRST symmetry transformations
($s_{ab}$) for {\it all} the fields that are present in the Lagrangian densities (1).

We end our present discussion with some remarks connected with  Sec. 3. as well as Sec. 4.
First, we note that we have {\it not} utilized  the  idea of HC  anywhere. Rather, we have used 
{\it only} the (anti-)BRST (i.e. quantum gauge) invariant restrictions on the superfields to derive the proper (anti-)BRST symmetry transformations in Sec. 3
and 4. Second, a close look at Eqs. (16), (22), (24) and (26) demonstrate that $s_b\leftrightarrow \partial_{\bar\theta}$
(i.e. the BRST symmetry transformation is  intimately   connected with the translational generator  $\partial_{\bar\theta}$
along the {\it anti-chiral} super-submanifold of the ($D$, 2)-dimensional supermanifold). By contrast,
 the anti-BRST symmetry transformation is deeply connected (i.e. $s_{ab}\leftrightarrow \partial_{\theta}$) with 
the translational generator $\partial_{\theta}$ along the $\theta$-direction of the {\it chiral}
super-submanifold of the ($D$, 2)-dimensional supermanifold. Finally, the ACSA to BRST formalism
 produces the (anti-)BRST  symmetry transformations for the gauge, (anti-)ghost and matter fields {\it together}
which is {\it not} the case with the application of HC {\it alone}.

\noindent
\section{Invariance of ${\cal L}_B$ and ${\cal L}_{\bar B}$ Under the (Anti-)BRST Symmetries: ACSA to BRST Formalism}

In this section, we capture the (anti-)BRST invariance of the Lagrangian densities
${\cal L}_B$ and ${\cal L}_{\bar B}$ (cf. Eq. (4)) in the language of ACSA to BRST 
formalism. Towards this objective in mind, we generalize, first of all, the Lagrangian density ${\cal L}_B$
to its counterpart {\it anti-chiral} super Lagrangian density $\tilde {\cal L}^{(ac)}_B$ (with $\tilde B^{(b)} (x,\bar\theta) = B(x)$) as:
\begin{eqnarray}
{\cal L}_B  &\longrightarrow & \tilde {\cal L}^{(ac)}_B  =  -\,\frac {1}{4}\, \tilde F^{\mu\nu( ac)}(x, \bar\theta)\cdot\,\tilde F_{\mu\nu}^{( ac)}(x, \bar\theta)\nonumber\\
& + & \bar\Psi^{(b)}(x, \bar\theta)\;( i\;\gamma^{\mu}\,\partial_{\mu} - m)\;\Psi^{(b)}(x, \bar\theta)\nonumber\\
& -&  \;\bar\Psi^{(b)}(x, \bar\theta)\;\gamma^{\mu}\; B_{\mu}^{(b)}(x, \bar\theta)\;\Psi^{(b)}(x, \bar\theta) + B(x)\cdot
\Bigl [\partial_{\mu} B^{\mu (b)}(x, \bar\theta)\Bigr ]\nonumber\\
& +&  \frac {1}{2}\;\Big[ B(x)\cdot B(x) + \tilde {\bar  B}^{(b)}(x, \bar\theta) \cdot\tilde {\bar  B}^{(b)}(x, \bar\theta)\Big]\nonumber\\
& - & i \;\partial_{\mu}\bar F^{(b)}(x, \bar\theta)\cdot\partial^{\mu} F^{(b)}(x, \bar\theta)\nonumber\\
 & +&  \partial_\mu\bar F^{(b)}(x,\bar\theta)\cdot
\Big[B^{\mu(b)}(x,\bar\theta)\times F^{(b)}(x,\bar\theta)\Big],
\end{eqnarray}
where {\it all} the symbols have been explained earlier in the super expansions (16), (22), (24) and  (26) except
$F_{\mu\nu}^{(ac)}(x,\bar\theta)$ which we explain, namely;  
\begin{eqnarray}
F ^{(ac)}_{\mu\nu}  & = & \partial_\mu B^{(b)}_\nu (x,\bar\theta)- \partial_\nu B^{(b)}_\mu (x,\bar\theta) + i\;\Big [B^{(b)}_\mu(x,\bar\theta)\times B^{(b)}_\nu(x,\bar\theta)\Big] \nonumber\\
& \equiv & F _{\mu\nu}(x) + \bar\theta\; (i\;F _{\mu\nu}\times C)\equiv F _{\mu\nu}(x) + \bar\theta\;(s_b\;F _{\mu\nu}(x)),
\end{eqnarray}
where the super expansion of $B^{(b)}_\mu(x,\bar\theta)$ is given in Eq. (24).
Substitutions of the super expansions for $\tilde   B^{(b)}(x, \bar\theta), F^{(b)}(x,\bar\theta), {\bar F}^{(b)}(x,\bar\theta)$
and $ F^{(ac)}_{\mu\nu}(x, \bar\theta)$ into the above super Lagrangian density lead to the following
\begin{eqnarray}
&&\tilde {\cal L}^{(ac)}_B = {\cal L}_B + \bar\theta\; \partial_\mu [ B\cdot D^\mu C],
\end{eqnarray}
modulo some other explicit total spacetime derivative terms ({\it without} being the coefficient of $\bar\theta$).
Thus, it is clear that we have the following: 
\begin{eqnarray}
&&\frac {\partial}{\partial\bar\theta} \;\tilde {\cal L}^{(ac)}_B = \partial_\mu 
\;[ B\cdot D^\mu C]\quad\Longleftrightarrow\quad s_b \;{\cal L}_B = \partial_\mu [ B\cdot D^\mu C].
\end{eqnarray}
The above quantity is {\it true} because of the identification of  BRST symmetry 
transformations $(s_b)$ with the translational generator ($\partial_{\bar\theta}$)
along the Grassmannian direction of the {\it anti-chiral} super-submanifold 
of the {\it general} ($D$, 2)-dimensional supermanifold.  
Geometrically, the super Lagrangian density $\tilde {\cal L}^{(ac)}_B$
is the {\it sum} of composite {\it anti-chiral} superfields (in addition to the 
ordinary fields) such that its translation along $\bar\theta$-direction
of the anti-chiral super-submanifold produces a {\it total}
spacetime derivative  (i.e $s_b \,{\cal L}_B = \partial_\mu [ B\cdot D^\mu C]$) in the {\it ordinary} spacetime thereby rendering the action integral 
($S = \int d^{D}x \; {\cal L}_B$) invariant.

To capture the anti-BRST invariance of the Lagrangian density ${\cal L}_{\bar B}$
(cf. Eq. (4)), we generalize the ordinary fields of {\it it} to the {\it chiral} super Lagrangian density $\tilde {\cal L}^{(c)}_{\bar B}$
(with input  $\tilde {\bar B}^{(ab)}(x,\theta) = \bar B(x)$) as:
\begin{eqnarray}
{\cal L}_{\bar B}& \longrightarrow & \tilde {\cal L}^{(c)}_{\bar B}  = 
 -\;\frac {1}{4}\; \tilde F^{\mu\nu( c)}(x, \theta)\cdot\tilde F_{\mu\nu}^{( c)}(x, \theta)\nonumber\\
& + & \bar\Psi^{(ab)}(x, \theta)\;( i\;\gamma^{\mu}\,\partial_{\mu} - m)\;\Psi^{(ab)}(x, \theta)\nonumber\\
& - & \;\bar\Psi^{(ab)}(x, \theta)\;\gamma^{\mu}\; B_{\mu}^{(ab)}(x, \theta)\;\Psi^{(ab)}(x, \theta) -  {\bar B}(x)\cdot
\Big[\partial_{\mu} B^{\mu (ab)}(x, \theta)\Big]\nonumber\\
& + & \frac {1}{2}\;\Big[\tilde B^{(ab)}(x, \theta)\cdot\tilde B^{(ab)}(x, \theta) +  {\bar  B}(x) \cdot {\bar  B}(x)\Big]\nonumber\\
 & - & i\; \partial_{\mu}\bar F^{(ab)}(x, \theta)\cdot\partial^{\mu} F^{(ab)}(x, \theta)\nonumber\\
 & + & 
\Big [B_{\mu}^{(ab)}(x,\theta)\times {\bar F}^{(ab)}(x,\theta))\Big]\cdot\partial^\mu F^{(ab)}(x,\theta),
\end{eqnarray}
where all the symbols have been explained earlier in the super expansions (29), (31), (32) and (33) except $F_{\mu\nu}^{( c)}(x, \theta)$
which we explicitly elaborate as follows
\begin{eqnarray}
F ^{(c)}_{\mu\nu}  & = & \partial_\mu B^{(ab)}_\nu (x,\theta)- \partial_\nu B^{(ab)}_\mu (x,\theta) 
+ i\;\Big[B^{(ab)}_\mu (x, \theta) \times B^{(ab)}_\nu(x,\theta)\Big] \nonumber\\
& \equiv & F _{\mu\nu}(x) + \theta\; (i\;F _{\mu\nu}\times \bar C)\equiv F _{\mu\nu}(x) + \theta\;(s_{ab}\;F _{\mu\nu}(x)),
\end{eqnarray}
where we have substituted the expansion for $ B_\mu^{(ab)}$ (cf. Eq. (31)).
The substitutions of all the {\it chiral} superfields with superscript $(ab)$ into (38) lead to the following:
\begin{eqnarray}
&&\tilde {\cal L}^{(c)}_{\bar B} = {\cal L}_{\bar B} + \theta\; \partial_\mu [ -\;\bar B\cdot D^\mu \bar C],
\end{eqnarray}
modulo some other explicit total spacetime derivative terms ({\it without} being the coefficient of  $\theta$).
Ultimately, we obtain the following mapping: 
\begin{eqnarray}
&&\frac {\partial}{\partial\theta} \;\tilde {\cal L}^{(c)}_{\bar B} = \partial_\mu 
\;[ -\; \bar B\cdot D^\mu \bar C]\quad\Longleftrightarrow \quad s_{ab} \;{\cal L}_{\bar B }= -\;\partial_\mu [ \bar B\cdot D^\mu\bar  C].
\end{eqnarray}
Geometrically, the above equation implies that the {\it chiral} super Lagrangian density is 
a very specific sum of the {\it chiral} superfields that have been obtained after the applications of 
anti-BRST invariant restrictions and some {\it ordinary} fields such that 
its translation along $\theta$-direction of the {\it chiral} super-submanifold produces a
total spacetime derivative (i.e. $ s_{ab} \,{\cal L}_{\bar B} = - \partial_\mu [ \bar B\cdot D^\mu \bar C]$) in the {\it ordinary} 
space thereby rendering the {\it ordinary} action integral 
($S = \int d^{D}x\;{\cal L}_{\bar B}$) invariant.

We can also capture the anti-BRST invariance of the Lagrangian density ${\cal L}_B$
and BRST invariance of the Lagrangian density  ${\cal L}_{\bar B}$ within the framework of 
ACSA to BRST formalism. In this context, we note that the Lagrangian density ${\cal L}_B$ can
 be generalized to the {\it chiral} super Lagrangian density $\tilde{\cal L}_B^{(c)}$ as:
\begin{eqnarray}
{\cal L}_B&\longrightarrow &\tilde {\cal L}^{(c)}_B = 
 -\;\frac {1}{4}\;\tilde F^{\mu\nu( c)}(x, \theta)\cdot \tilde F_{\mu\nu}^{( c)}(x, \theta)\nonumber\\
& + & \bar\Psi^{(ab)}(x, \theta)\;( i\;\gamma^{\mu}\,\partial_{\mu} - m)\;\Psi^{(ab)}(x, \theta)\nonumber\\
& - & \;\bar\Psi^{(ab)}(x, \theta)\;\gamma^{\mu}\; B_{\mu}^{(ab)}(x, \theta)\;\Psi^{(ab)}(x, \theta) + \tilde B^{(ab)}(x, \theta))\cdot
\Big[\partial_{\mu} B^{\mu (ab)}(x, \theta)\Big]\nonumber\\
& + & \frac {1}{2}\;\Big[\tilde B^{(ab)}(x)\cdot \tilde B^{(ab)}(x) +  {\bar  B}(x) \cdot {\bar  B}(x)\Big]
 -  i\; \partial_{\mu}\bar F^{(ab)}(x, \theta)\cdot\partial^{\mu} F^{(ab)}(x, \theta)\nonumber\\
 & + &\partial_\mu \bar F^{(ab)}(x,\theta)\cdot
\Big[B^{(ab)\mu}(x,\theta)\times F^{(ab)}(x,\theta)\Big],
\end{eqnarray}
where all the symbols have been explained in Sec. 4 and in present section.
It is straightforward to check that the substitutions of all these super expansions 
into the above super Lagrangian density lead to the following explicit result, namely; 
\begin{eqnarray}
\tilde {\cal L}^{(c)}_B &=& {\cal L}_B + \theta\;\Big [-\;\partial_\mu \Big \{\bar B+ (C\times\bar C)\big\}\cdot \partial^\mu\bar C
+ (B + \bar B + (C\times\bar C))\cdot D_\mu \partial^\mu\bar C\Big ]\nonumber\\
&\equiv & {\cal L}_B  + \theta\; \partial_\mu \,[B\cdot \partial^\mu \bar C],
\end{eqnarray}
where the final/last expression 
has been obtained after the application of CF-condition: $B + \bar B + (C\times\bar C) = 0$.
Now, it is crystal clear that we have the following mapping between the Grassmannian partial derivative of {\it chiral}
super submanifold and anti-BRST symmetry transfromation
$s_{ab}$ in the {\it ordinary} space:
\begin{eqnarray}
&&\frac {\partial}{\partial\theta} \;\tilde {\cal L}^{(c)}_B = \partial_\mu 
\;[ B\cdot \partial^\mu \bar C]\quad\Longleftrightarrow\quad s_{ab} \;{\cal L}_B = \partial_\mu [ B\cdot \partial^\mu\bar  C],
\end{eqnarray}
which completely agrees with our observation in Eq. (5). Thus, we note that the Lagrangian density ${\cal L}_B$
{\it also} respects the anti-BRST symmetry transformations on a constrained hypersurface\footnote{
It is obvious that, if the CF-condition is taken into account, the action
integral $S = \int\, d^{D} x \,{\cal L}_B$ remains invariant under the {\it anti-BRST}
symmetry transformations $s_{ab}$, too.} in the flat $D$-dimensional Minkowskian
sapcetime manifold (which is defined by the CF-condition:  $B + \bar B + (C\times\bar C) = 0$). It is  interesting to mention,
in passing, that we {\it also} have the absolutely anticommuting  $s_{(a)b}$ on {\it this}
 hypersurface where $B + \bar B + (C\times\bar C) = 0$ is true.

At this stage, we dwell a bit on the BRST invariance of the Lagrangian density  ${\cal L}_{\bar B}$.
Towards this objective in mind, first of all, we generalize {\it this} Lagrangian density to the super
{\it anti-chiral} Lagrangian density ${\cal L}^{(ac)}_{\bar B}$ as
\begin{eqnarray}
{\cal L}_{\bar B}&\longrightarrow & \tilde {\cal L}^{(ac)}_{\bar B}  =  -\;\frac {1}{4}\; \tilde F^{\mu\nu( ac)}(x, \bar\theta)\cdot\;\tilde F_{\mu\nu}^{( ac)}(x, \bar\theta)\nonumber\\
& + & \bar\Psi^{(b)}(x, \bar\theta)\;( i\;\gamma^{\mu}\,\partial_{\mu} - m)\;\Psi^{(b)}(x, \bar\theta)\nonumber\\
& - & \;\bar\Psi^{(b)}(x, \bar\theta)\;\gamma^{\mu}\; B_{\mu}^{(b)}(x, \bar\theta)\;\Psi^{(b)}(x, \bar\theta) -\; \tilde{\bar B}^{(b)}(x, \theta)\cdot
\Big[\partial_{\mu} B^{\mu (b)}(x, \bar\theta)\Big]\nonumber\\
& + & \frac {1}{2}\;\Big [ B(x)\cdot B(x) + \tilde {\bar  B}^{(b)}(x, \bar\theta) \cdot\tilde {\bar  B}^{(b)}(x, \bar\theta)\Big]\nonumber\\
& - & i \;\partial_{\mu}\bar F^{(b)}(x, \bar\theta)\cdot\partial^{\mu} F^{(b)}(x, \bar\theta)\nonumber\\
 & + & 
\Big[B^{\mu(b)}(x,\bar\theta)\times \bar F^{(b)}(x,\bar\theta)\Big]\cdot\partial_\mu F^{(b)}(x,\bar\theta),
\end{eqnarray}
where all the symbols/notations have been explained earlier in Sec. 3 and in the present section.
We note that the above super Lagrangian density is the  {\it sum} of the composite superfields (derived after
 the application of BRST invariant restrictions) {\it and} ordinary fields. We are now in the position 
 to operate a derivative w.r.t. $\bar\theta$ on the above {\it super} Lagrangian density as 
\begin{eqnarray}
&&\frac {\partial}{\partial\bar\theta} \;\tilde {\cal L}^{(ac)}_{\bar B} = \partial_\mu 
\;[-\; \bar B\cdot \partial^\mu C]~~~~~\Longleftrightarrow ~~~s_b \;{\cal L}_{\bar B} = \partial_\mu [-\; \bar B\cdot \partial^\mu C],
\end{eqnarray}
 where the  mapping between the {\it superspace} and  {\it ordinary} space has been taken into account.
 In fact, the above result has been obtained due to the fact that the super {\it anti-chiral} Lagrangian
 density (45) can be written in the following explicit form 
\begin{eqnarray}
\tilde {\cal L}^{(ac)}_{\bar B} &=& {\cal L}_{\bar B} + \bar \theta\;\Big [\partial_\mu \big \{ B+ (C\times\bar C)\big\}\cdot \partial^\mu C
-(B + \bar B + (C\times\bar C))\cdot D_\mu \partial^\mu C\Big ]\nonumber\\
&\equiv & {\cal L}_{\bar B}  + \bar\theta\; \partial_\mu \,[-\;\bar B\cdot \partial^\mu  C],
\end{eqnarray}
 where the {\it final} expression has been obtained after the application  of  CF-condition: $B + \bar B + (C\times\bar C) = 0$.
 It is worthwhile to mention here that the result of (46) has been obtained after the derivation of (47) and the application of the derivative
 w.r.t. $\bar\theta$ on it (i.e. Eq. (47)). This statement is {\it true} because of our observations in Eq. (5) and Eqs. (37), (41), (44)
 and (46). In other words, there is a {\it precise} agreement between the results obtained in the {\it ordinary} space and {\it superspace}
(with the help of the mappings: $s_b \leftrightarrow \partial_{\bar\theta}, s_{ab} \leftrightarrow \partial_{\theta}$).

 We end this section with the {\it final} remark that
 we have {\it already } captured the essential features of the (anti-)BRST invariance of the Lagrangian densities 
${\cal L}_B$ and ${\cal L}_{\bar B}$ within the framework of ACSA to BRST formalism.

\section{Nilpotency and Absolute Anticommutativity  Properties of the Fermionic Conserved (Anti-)BRST Charges: ACSA}

\vskip .5cm

\noindent

In this section, we capture the off-shell nilpotency and absolute anticommutativity
 properties of the (anti-)BRST charges within the framework of ACSA to BRST formalism.
In the proof of absolute anticommutativity property, we invoke the CF-condition at 
appropriate places. At the very onset, we would like to lay  emphasis on the fact that 
our knowledge of the {\it ordinary} {\it space } BRST formalism and its connection with 
the superspace/superfield approach to BRST formulation has helped us in {\it all} our theoretical 
discussions of this section. In other words, our understandings of BRST formalism in {\it both} 
the {\it spaces} is intertwined together in a beautiful and useful manner. It is because of this 
reason that we have been able to express the mathematical properties of $Q_{(a)b}$ in the 
language of ACSA.

Towards our main discussions, first of all, we discuss the {\it nilpotency} property of the  conserved (anti-)BRST 
charges within the framework of ACSA to BRST formalism. It is straightforward to check that the 
above conserved charges (i.e. nilpotent (anti-)BRST charges) can be written as (cf. Eq. (12)). 
\begin{eqnarray}
Q_{ab} &=& \frac {\partial}{\partial\theta}\int\; d^{D-1}x\;\Big[ i\;\bar F^{(ab)}(x,\theta)\;\cdot\dot F^{(ab)}(x,\theta)  
- i\;\bar B(x)\cdot B_0^{(ab)}(x, \theta)\Big]\nonumber\\
&\equiv & \int\; d\theta\;\int\; d^{D-1}x\;\Big[ i\;\bar F^{(ab)}(x,\theta)\;\cdot\dot F^{(ab)}(x,\theta)  
- i\;\bar B(x)\cdot B_0^{(ab)} (x,\theta)\Big],\nonumber\\
 Q_{b} &=& \frac {\partial}{\partial\bar\theta}\int\; d^{D-1}x\;\Big[B(x)\cdot B_0^{(b)}(x,\bar\theta) + 
 i\;\dot {\bar F}^{(b)}(x,\bar\theta)\;\cdot F^{(b)}(x,\bar\theta)  \Big]\nonumber\\
&\equiv & \int\; d\bar\theta\;\int\; d^{D-1}x\;\Big[B(x)\cdot B_0^{(b)}(x,\bar\theta) + 
 i\;\dot {\bar F}^{(b)}(x,\bar\theta)\;\cdot F^{(b)}(x,\bar\theta)\Big],
\end{eqnarray}
where we have taken $\tilde {\bar B}^{(ab)}(x,\theta) = \bar B(x)$ and $\tilde B^{(b)}(x,\bar\theta) = B(x)$
which have been derived earlier (primarily due to: $s_{ab}\,\bar B(x) = 0, s_b\, B(x)  = 0$). Rest of {\it all} the symbols
 have been explained earlier. We have also established that: $s_b\longleftrightarrow \partial_{\bar\theta}$ and  
 $s_{ab}\longleftrightarrow \partial_\theta$. Thus, the above expressions for the (anti-)BRST charges (in the 
{\it superspace} formulation) can be translated into the {\it ordinary} space formulation  in the language of the nilpotent (anti-)BRST transformations
and {\it ordinary} fields as:
\begin{eqnarray}
&&Q_{ab} = s_{ab} \int d^{D-1}x\;\Big[i\,\bar C(x)\,\dot C(x) - \,\bar B(x)\cdot A_0 (x)\Big],\nonumber\\
&&Q_b = s_b \int d^{D-1}x\;\Big[B(x)\cdot A_0(x) + i\,\dot{\bar C}(x)\cdot C(x)\Big]. 
\end{eqnarray}
It is now crystal clear that we have:
\begin{eqnarray}
&&\partial_\theta \;Q_{ab} = 0\; \Longleftrightarrow \, \partial_\theta^2 = 0,~~\, s_{ab}\;Q_{ab} = -\,i\,{\{Q_{ab}, Q_{ab}}\} = 0\, \Longleftrightarrow\, s_{ab}^2 = 0,\nonumber\\
&&\partial_{\bar\theta}\; Q_b = 0 \;\,~\Longleftrightarrow \, \partial_{\bar\theta}^2 = 0,~~\;\, s_b\;Q_b \;\;= -\,i\,{\{Q_b, Q_b}\} = 0\;\;\;\, \Longleftrightarrow \, s_b^2 = 0. 
\end{eqnarray}
In other words, the nilpotency of the (anti-)BRST charges (i.e. $Q_{(a)b}^2 = 0)$ is deeply connected with the nilpotency 
($\partial_\theta^2 = \partial_{\bar\theta}^2 = 0$) of the translational generators $(\partial_\theta, \partial_{\bar\theta})$
along Grssmannian directions
as well as the nilpotency ($s_{(a)b}^2 = 0$) of the (anti-) BRST symmetry transformations $(s_{(a)b})$.

As far as the proof of absolute anticommutativity is concerned, we begin with such a proof {\it first}
in the {\it ordinary} space\footnote {We purposefully perform this exercise to 
demonstrate that our knowledge in the {\it ordinary} space and {\it superspace} is intertwined 
(for all the discussions contained in this section).}
 by exploiting the beauty and strength of the symmetry principles (i.e. the continuous symmetries 
and their generators). In view of this, first of all, we recast the expression for the BRST charge $Q_b$
(cf. Eq. (12)) in an appropriate form by using the CF-condition $B + \bar B + (C\times\bar C) = 0$.
This suitable (i.e. modified $but$ equivalent) form of $Q_b$ is as follows
\begin{eqnarray}
Q_b & = & \int d^{D-1}x\;\Big[\dot{\bar B}\cdot C - \bar B\cdot D_0 C - (C\times\bar C)\cdot D_0 C \nonumber\\ 
& + &\frac {1}{2}  \;\dot {\bar C}\cdot (C\times\bar C) + (\dot C\times\bar C)\cdot C\Big]\nonumber\\
& \equiv & \int d^{D-1}x\;\Big[\dot{\bar B}\cdot C - \bar B\cdot D_0 C + \frac {1}{2}\; \dot {\bar C}\cdot (C\times C)\nonumber\\
& - & \;i\;(C\times\bar C)\cdot(A_0 \times C)\Big],      
\end{eqnarray}
where we have used $D_0 C = \partial_0 C + i\,(A_0\times C)\equiv \dot C + i\;(A_0\times C)$. The above 
{\it final} expression of $Q_b$ can be written as an {\it anti-BRST} {\it exact} quantity:
\begin{eqnarray}
&&Q_b = s_{ab}\;\int d^{D-1}x\;\Big[i\,C\cdot\dot C - \frac {1}{2}\; C\cdot (A_0 \times C)\Big].
\end{eqnarray}
The expression for $Q_b$, in the above form, proves the absolute anticommutativity of the (anti-)BRST
charges in the following manner
\begin{eqnarray}
&&s_{ab}Q_b = -\,i\;{\{Q_b, Q_{ab}}\} = 0\quad\Longleftrightarrow \quad s_{ab}^2 = 0,
\end{eqnarray}
where we have used the idea of continuous symmetry generator and the off-shell nilpotency of the 
anti-BRST symmetry transformation ($s_{ab}$). Taking into account our knowledge of $s_{ab}\longleftrightarrow \partial_\theta$, 
the above expression (52) can be written, within the framework of ACSA to BRST formalism, as: 
\begin{eqnarray}
 Q_b     & = & \frac {\partial}{\partial\theta}\;\Big [ \int\; d^{D-1}x\;\Big\{i\; F^{(b)}(x,\bar\theta)\cdot \dot F^{(b)}(x,\bar\theta)\nonumber\\
 & - &  \frac {1}{2} \;F^{(b)}(x,\bar\theta)\cdot \Big[B^{(b)}_0(x,\bar\theta)\times F^{(b)}(x,\bar\theta)\Big]\Big\}\Big]\nonumber\\
  & \equiv  &   \int\; d\theta\;\int\; d^{D-1}x\;\Big [ \Big\{i\; F^{(b)}(x,\bar\theta)\cdot \dot F^{(b)}(x,\bar\theta)\nonumber\\
 & - &  \frac {1}{2} \;F^{(b)}(x,\bar\theta)\cdot \Big[B^{(b)}_0(x,\bar\theta)\times F^{(b)}(x,\bar\theta)\Big]\Big\}\Big].
\end{eqnarray}
From the above expression, it is clear that:
\begin{eqnarray}
&&\partial_\theta Q_b  = 0\quad\Longleftrightarrow\quad \partial_\theta^2 = 0\quad\Longleftrightarrow \quad s_{ab}Q_b = - i\;{\{Q_b, Q_{ab}}\} = 0.
\end{eqnarray}
Thus, we note that the absolute anticommutativity of the BRST charge $(Q_b)$ {\it with} the anti-BRST charge $(Q_{ab})$ 
is deeply connected with the nilpotency $(\partial_\theta^2 = 0)$ of the  translational generator $(\partial_\theta)$ along
the $\theta$-direction of the {\it chiral} super-submanifold of the general ($D$, 2)-dimensional
supermanifold (on which the fields of our ordinary  $D$-dimensional gauge  theory have been  generalized to superfields).

Against the above discussions  as the backdrop, we prove that the absolute anticommutativity
of the anti-BRST charge $(Q_{ab})$ {\it with} the BRST charge $(Q_b)$ is connected with the 
nilpotency of the translational generators   ($\partial_{\bar\theta}$) along the $\bar\theta $-direction 
of the {\it anti-chiral} super-submanifold (of the ($D$, 2)-dimensional general supermanifold).
To accomplish this goal, we note that the anti-BRST charge can be written, in the {\it BRST-exact} form,
as
\begin{eqnarray}
&&Q_{ab} = s_b \;\int d^{D-1}x\;\Big[-\;i\;{\bar C}\cdot \dot{\bar C} + \frac {1}{2}\; {\bar C} \cdot (A_0\times \bar C)\Big],
\end{eqnarray}
where we have already used the CF-condition $B+\bar B+ (C\times\bar C) = 0$ to 
recast the expression for $Q_{ab}$ in an appropriate form\footnote{The algebraic computations are on exactly {\it similar} lines as we have done for the proof 
of the absolute anticommutativity of BRST charge ($Q_b$) {\it with} the anti-BRST charge $(Q_{ab}$).} so that it could be written as 
(56). Within the framework of ACSA to 
BRST formalism, the above form (56) can be written (on the  ($D$, 1)-dimensional {\it anti-chiral} super-submanifold) as:
\begin{eqnarray}
Q_{ab}     & = & \frac {\partial}{\partial\bar\theta}\;\Big [ \int\; d^{D-1}x\;\Big\{-i\;\;\bar  F^{(b)}(x,\bar\theta)\cdot \dot {\bar F}^{(b)}(x,\bar\theta)\nonumber\\
 & + &  \frac {1}{2}\;\bar F^{(b)}(x,\bar\theta)\cdot (B^{(b)}_0(x,\bar\theta)\times {\bar F}^{(b)}(x,\bar\theta))\Big\}\Big]\nonumber\\
 & \equiv  &   \int\; d\theta\; \int\; d^{D-1}x\;\Big [-i\;\;\bar  F^{(b)}(x,\bar\theta)\cdot \dot {\bar F}^{(b)}(x,\bar\theta)\nonumber\\
 & + &  \frac {1}{2}\;\bar F^{(b)}(x,\bar\theta)\cdot (B^{(b)}_0(x,\bar\theta)\times {\bar F}^{(b)}(x,\bar\theta))\Big].
\end{eqnarray}
Thus, it is straightforward to note that we have: 
\begin{eqnarray}
&&\partial_{\bar\theta} Q_{ab} = 0\,\Longleftrightarrow\,\partial_{\bar\theta}^2 = 0,\,
s_b  Q_{ab} = - \;i\;{\{Q_{ab}, Q_b}\} = 0\,\Longleftrightarrow \, s_b^2 = 0.
\end{eqnarray}
Finally, we remark that the absolute anticommutativity of the anti-BRST  charge {\it with} the BRST charge 
is connected with the nilpotency of the translational generator along $\bar\theta$-direction of the ($D$, 1)-dimensional 
{\it anti-chiral} super-submanifold (of the general ($D$, 2)-dimensional supermanifold). This observation,
in turn, implies that the above absolute anticommutativity is also deeply connected with the nilpotency
$(s_b^2 = 0)$ of the BRST symmetry transformations $(s_b)$.

\section{Conclusions}

\vskip .5cm

In our present investigation, we have exploited the theoretical strength  of  ACSA to BRST formalism
to derive the (anti-)BRST symmetry transformations by demanding that the (anti-)BRST (i.e. quantum gauge) 
invariant quantities  {\it must}
 be independent of the ``soul" coordinates. In terms of the geometrical quantities defined on the 
(anti-)chiral super-submanifolds {\it and} (anti-)chiral  superfields (derived after the application of the (anti-)BRST 
invariant restrictions), we have been able to express the conserved (anti-)BRST charges of our theory
in the language of ACSA to BRST formalism.
This exercise, in turn, has helped us to capture the properties of the off-shell nilpotency and absolute anticommutativity
of the conserved charges of our {\it interacting} $D$-dimensional non-Abelian 1-form gauge theory.

One of the {\it novel} observations of our present endeavor is the proof of  absolute anticommutativity
of the (anti-)BRST conserved charges {\it despite} the fact that we have considered {\it only}  the 
(anti-)chiral super expansions of the (anti-)chiral superfields within the framework of ACSA to BRST formalism. In this proof, the celebrated CF-condition [17]
has played a crucial role. In fact, our knowledge of  various key aspects of the superfield approach to BRST formalism (on the suitably chosen
supermanifolds)
and their deep connection with the nilpotent (anti-)BRST symmetries in the {\it ordinary} space has helped us in accomplishing  the above goal 
(which is one of the highlights of our present investigation). We have been able to express the nilpotency 
and absolute anticommutativity  properties in the {\it ordinary} space, too. However, this has been possible  
 because of our  deep understanding of  various aspects of the superfield approach to BRST formalism .

Within the framework of ACSA to BRST formalism, the observation of absolute anticommutativity  of the (anti-)BRST
charges is a completely {\it novel} result because we have studied various ${\cal N} = 2$ SUSY quantum
mechanical models and applied the (anti-)chiral supervariable approach to derive the nilpotent ${\cal N} = 2$ 
supersymmetric transformations  {\it but} the corresponding ${\cal N} = 2$ SUSY charges have been shown to be 
{\it not}\footnote{
We take a simple example of the ${\cal N} = 2$ SUSY quantum mechanical model of a 1$D$ harmonic oscillator to corroborate this statement in our Appendix B.}
absolutely anticommuting in nature [23-27]. In fact, it has been shown that the anticommutator of 
${\cal N} = 2$ SUSY charges generates  the time translation of the variable on which 
it operates. In other words, the anticommutator of the ${\cal N} = 2$ super charges leads to the 
derivation of Hamiltonian for the ${\cal N} = 2$ SUSY quantum mechanical model. Against this backdrop,
it is clear that the observation of the absolute anticommutativity property of the (anti-)BRST charges is 
a completely {\it novel} result within the framework of ACSA to BRST formalism. We discuss {\it briefly} about this
surprisingly {\it novel} result in our Appendix A.

The ideas of ACSA to BRST formalism are simple and straightforward and they lead to the
derivation of (anti-)BRST symmetries for {\it all }the fields {\it together}. This should 
be contrasted  with the  HC which leads to the derivation of the off-shell nilpotent (anti-)BRST symmetries for the gauge and (anti-)ghost 
fields {\it only}. We plan to extend our ideas in the context  of discussions for the higher $p$-form 
($p$ =  2, 3, 4,...) gauge theories (within the framework of (anti-)chiral  superfield approach to BRST formalism) so that the  ACSA to BRST 
formalism
could  be firmly established. In this context, it is gratifying  to mention that we have already 
applied our ideas of ACSA to BRST formalism in the case of  a {\it free} 4$D$  Abelian 2-form gauge theory and have proven the  
 absolute anticommutativity of the (anti-)BRST charges (see, e.g. [15] for details). In this proof, 
we have shown that the CF-type restriction (for the Abelian 2-form gauge theory) plays a decisive role.\\

\noindent
{\bf Acknowledgements}\\

\vskip .5cm

\noindent
The present investigation has been carried out under the BHU-fellowship received  by S. Kumar and the
DST-INSPIRE fellowship (Govt. of India) awarded to B. Chauhan. Both these authors express their gratefulness to
the above local and national funding agencies, respectively,  for their financial supports.\\

\begin{center}
{\bf Appendix A:   Absolute Anticommutativity and Full Super Expansion}\\
\end{center}
\noindent

\vskip .5cm

\noindent
To highlight the {\it novel} observation of the absolute anticommutativity of the (anti-)BRST charges (in our
present investigation), we discuss here concisely the connection of {\it this} property with the {\it full} super 
expansion of the superfields along {\it all} the Grassmannian direction, of the ($D$, 2)- dimensional
supermanifold on which the  fields of a given $D$-dimensional {\it ordinary} gauge theory are generalized. As we know,
one of the decisive features of BRST and anti-BRST symmetry transformations is the absolute anticommutativity
property which primarily captures the linear {\it independence} of these symmetry transformations. Expressed in terms of the 
translational generators ($\partial_{\bar\theta}, \partial_\theta$) along the Grassmannian directions
of the ($D$, 2)-dimensional supermanifold, we observe the following (with inputs $\partial_{\bar\theta} \longleftrightarrow s_b ,
\partial_\theta \longleftrightarrow s_{ab}$):
\[
s_b\,s_{ab} + s_{ab}\,s_b = 0 \qquad\quad\Longleftrightarrow \qquad\quad\partial_{\bar\theta}\,\partial_\theta  + \partial_\theta 
\partial_{\bar\theta} = 0.
\eqno(A.1)\]
Since the continuous symmetry transformations $s_{(a)b}$ are deeply connected with (and generated by) the Noether conserved
charges $Q_{(a)b}$, the above relationship  (A.1) can be {\it also} translated into the following:
\[
 Q_b\;Q_{ab} + Q_{ab}\;Q_b = 0\qquad\quad\Longleftrightarrow \qquad\quad \partial_{\bar\theta}\,\partial_\theta  + \partial_\theta 
\partial_{\bar\theta} = 0.
\eqno(A.2)\]
We claim, in this Appendix, that (A.1) and (A.2) become quite obvious  and transparent 
when we take the {\it full} super expansions of the superfields defined on the ($D$, 2)-dimensional supermanifold.
However, this is {\it not} the case when we take {\it only} the (anti-)chiral super expansions of the (anti-)chiral 
superfields that are defined on the ($D$, 1)-dimensional super-submanifolds of the general ($D$, 2)-dimensional 
supermanifold (on which a given $D$-dimensional  {\it ordinary} gauge theory is generalized).
For instance, we have applied the (anti-)chiral supervariable approach to ${\cal N} = 2$ SUSY quantum mechanical models where 
the property of absolute anticommutativity is {\it not} satisfied (see, e.g. [19-22] for details).

To corroborate   the above statement, we begin with a generic superfield $\Omega\,(x, \theta, \bar\theta)$
which has  the following {\it full} super expansion  along {\it all} the Grassmannian directions 
of the ($D$, 2)-dimensional supermanifold, namely;
\[
\Omega \,(x, \theta, \bar\theta) = \omega\, (x) + \theta\, \bar P (x) + \bar\theta\, P(x) + i\;\theta\,\bar\theta\, Q(x),
\eqno (A.3)\]
where  $\omega \,(x)$, on the r.h.s., is the {\it basic} field of the $D$-dimensional gauge theory and the set $(P(x), \bar P(x), Q(x))$
represents the existence of {\it secondary} fields. The fermionic ($\theta^2 = \bar\theta^2 = 0, \theta\,\bar\theta
+ \bar\theta\,\theta = 0$) nature of the Grassmannian variables $(\theta,\,\bar\theta)$
demonstrate that, if $\Omega\, (x, \theta, \bar\theta)$ were fermionic in nature, the pair $(P(x), \bar P(x))$
would be bosonic and $Q(x)$ would be fermionic. On the other hand, if $\Omega\, (x, \theta, \bar\theta)$ 
were {\it bosonic}, the pair $(P(x), \bar P(x))$ would be {\it fermionic} and $Q(x)$ would be {\it bosonic}. It is elementary 
to check that the following are true for the super expansion (A.3), namely;
\[\frac {\partial}{\partial\theta}\,\frac {\partial}{\partial\bar\theta}\, \Omega \,(x, \theta, \bar\theta) = -\,i\,Q(x),
\qquad\qquad  \frac {\partial}{\partial\bar\theta}\,\frac {\partial}{\partial\theta} \, \Omega \,(x, \theta, \bar\theta) = + \,i \,Q(x).
\eqno (A.4)\]
The above relationship automatically establishes the following
\[
\Big(\frac {\partial}{\partial\theta}\,\frac {\partial}{\partial\bar\theta} + \frac {\partial}{\partial\bar\theta}\,
\frac {\partial}{\partial\theta}\Big)\,\Omega \,(x, \theta, \bar\theta) = 0,
\eqno (A.5)\]
which leads to the operator form of the relationship: $\partial_\theta\,\partial_{\bar\theta} + \partial_{\bar\theta}\partial_\theta = 0$.
This relationship, when translated into the ordinary space, leads to the connections that have been expressed in (A.1) and 
(A.2). Thus, the property of absolute anticommutativity, at the level of symmetry operators and conserved charges,  becomes
 very {\it natural}, automatic and transparent when we consider the {\it full}
super expansions of the superfields defined on the ($D$, 2)-dimensional supermanifold.
It is crystal clear that when we take {\it only} the (anti-)chiral super expansions  of the superfields,
the above relationship (A.4) and (A.5) {\it do} not become transparent and obvious.

In our present investigation, we have taken {\it only} the truncated  version of the super expansion (A.3). 
In other words, we have considered {\it only} the (anti-)chiral version of (A.3). Thus, the absolute 
anticommutativity ($Q_b\,Q_{ab} + Q_{ab}\,Q_b = 0$) of the charges $Q_{(a)b}$ is {\it not} obvious.
In fact, for the ${\cal N} = 2$ SUSY quantum mechanical models, it has been shown [19-22] that the conserved
nilpotent {\it super} charges do {\it not } absolutely anticommute within the framework of (anti-)chiral
supervariable approach to these models (see, Appendix B). Thus, the observation of absolute 
anticommutativity of the (anti-)BRST charges, within the framework of ACSA to BRST formalism, is a
completely {\it novel} result. Now, with the back up from  our 
earlier works [13-16, 18] and very recent  work [28],  we have been able to establish that the absolute anticommutativity 
of the (anti-)BRST charges is a {\it universal} truth when we apply the  ACSA to BRST formalism
in the cases of $p$-form $(p = 1, 2, 3,...)$ gauge theories as well as the reparameterization invariant theories
[28]. \\

\begin{center}
{\bf Appendix B:  ${\cal N} = 2$ SUSY QM Model of a  Harmonic Oscillator and Its  Nilpotent and Conserved Charges}\\
\end{center}

\vskip .5cm

\noindent
In this Appendix, we discuss, in a bit elaborate fashion,  our earlier work [19] on the ${\cal N} = 2$ SUSY QM system of a harmonic oscillator. However,
our present discussions are somewhat {\it  different} from [19]. This very interesting model  is described
by the following Lagrangian (with mass $m = 1$ and natural frequency $\omega$), namely;
\[
L_{0} = \frac {1}{2} \; {\dot x}^2 - \frac {1}{2} \; \omega^2 x^2 + i \; \bar \psi \;\dot \psi - \omega\; \bar \psi \;\psi, \eqno (B.1)\]
where $\dot x = ({dx}/{dt})$ and $\dot \psi = ({d\psi}/{dt}) $  are the ``generalized" velocities for the bosonic variable $x$ and its fermionic SUSY
counterpart $\psi$ w.r.t. the evolution parameter $t$ that characterizes the trajectory of the ${\cal N} = 2$ toy model of a 1$D$ SUSY harmonic  oscillator. In fact, we have a pair of fermionic variables  $\psi$  and $\bar\psi$   (i.e. ${\cal N} = 2$ SUSY partners) 
which obey: $\psi^2 = 0, \bar \psi^2 = 0,  \psi \bar \psi + \bar \psi \psi = 0$. The above Lagrangian $(B.1)$ respects 
($ s_1 L_{0}  = \frac {d}{dt}(\omega\,x\,\psi), \; s_2 L_{0}  = \frac {d}{dt}(-\,i\,\bar\psi \, \dot x)$) 
the following  {\it two} ${\cal N} = 2$ SUSY transformations (see, e.g. [19] for details)
\[s_1 x = i\;  \psi,\quad s_1 \psi = 0,\quad s_1 \bar \psi 
= -\,(\dot x + i\;\omega \;x ),\]
\[s_2 x = -\,i\;  \bar \psi,\quad s_2 \bar \psi = 0,\quad s_2 \psi 
= (\dot x - i\;\omega\; x ),\eqno (B.2)\] 
where $s_1$ and $s_2$ are nilpotent ($s_1^2 = 0, \;s_2^2 = 0$) of order {\it two} provided we use the EL-EOM ($\dot \psi + i\; \omega \psi = 0$ and $\dot {\bar \psi} - i\; \omega \bar \psi = 0$) {\it but } they do {\it not} absolutely anticommute (i.e. $s_1\,s_2 + s_2\,s_1 \neq 0$)
even if we use the {\it on-shell} conditions (i.e. EL-EOM).
The above continuous symmetries, according to Noether's theorem, lead to the derivation of the following  conserved ${\cal N} = 2$ SUSY charges $Q$ and $\bar Q$ 
\[Q =  (i\, p - \omega x) \; \psi \equiv (i\, \dot x - \omega x)\;\psi,\] 
\[\bar Q = -\, \bar \psi \;(i\, p + \omega x) \equiv  -\, \bar \psi \;(i\, \dot x + \omega x),\eqno (B.3)\]
where $p = \dot x$ is the momentum corresponding to the bosonic variable $x$. Using the following EL-EOMs, derived from the Lagrangian  $L_0$, namely;
\[\ddot x + \omega ^2\, x = 0,\qquad\;\;\ddot \psi + \omega ^2\, \psi = 0,\qquad\;\;\ddot {\bar\psi} + \omega ^2\, \bar\psi = 0,\]
\[\dot \psi + i\, \omega \, \psi = 0, \qquad\;\;\;\;\dot {\bar \psi} - i\, \omega \,\bar\psi = 0, \eqno (B.4)\]
it is straightforward to check that {\it both} the above ${\cal N} = 2$ SUSY charges are conserved (i.e. $\dot Q = 0, \;\dot{ \bar Q} = 0)$.
In other words, these charges remain {\it constant} w.r.t. the evolution parameter $t$ which parametrizes the trajectory of an ${\cal N} = 2$ 
SUSY QM model of a 1$D$ harmonic oscillator.

These conserved charges are {\it  also} the generators for the ${\cal N} = 2$ SUSY transformations $s_1$ and $s_2$ because 
it can be checked that the following   relationships lead to the derivations of the symmetry transformations $(B.2)$, namely;
\[s_1 \Phi = -\,i\; [\Phi, \; Q]_{\pm}, \quad \qquad s_2 \Phi = -\, i\; [\Phi, \; \bar Q]_{\pm}, \eqno (B.5)\] 
where $\Phi = x, \psi, \bar \psi$ is the generic variable of our theory and ($\pm$) signs, as the superscripts on 
the square bracket, denote the (anti)commutator for the generic variable
$\Phi$ being fermionic/bosonic in nature. The basic canonical brackets: $[x, p] = 1, \, i\, \{ \psi, \bar\psi\} = 1$ have to be exploited  in the verification of 
$(B.5)$  as far as the derivations of the ${\cal N} = 2$ SUSY transformations $s_1$ and $s_2$ are concerned. The above charges 
$Q$ and  ${\bar Q}$ are nilpotent ($ Q^2 = {\bar Q}^2 = 0)$ of order two because 
\[s_1 Q = -\,i\, \{Q, Q\}  = 0 \Rightarrow Q^2 = 0, \qquad s_2 \bar Q = -\,i\, \{\bar Q, \bar Q\} = 0 \Rightarrow \bar Q^2 = 0,\eqno (B.6)\]
where the l.h.s. can be computed  {\it easily} by the  direct applications of $s_1$ and $s_2$ on the charges $(B.3)$. The nilpotency property $(B.6)$
can {\it also} be proven by the following very useful and interesting  observations
\[ Q = s_1 \,(-\,i\,\bar\psi\,\psi), \qquad   \bar Q = s_2 \,(i\,\bar\psi\,\psi), \eqno (B.7)\]
which demonstrate that $s_1 \,Q = 0,\; s_2 \,\bar Q = 0$ due to the nilpotency property: $s_1 ^2 = s_2 ^2 = 0 $.
In other words, we find that the nilpotency (${s_1}^2 = {s_2}^2 = 0$) of $s_1$ and $s_2$  are deeply connected with the nilpotency ($Q^2 = {\bar Q}^2 = 0$) of $Q$ and $\bar Q$. It is worthwhile  to mention here that the expressions  for $Q$ and $\bar Q$ (cf. ($B.3$)) can {\it never} ever be expressed as some kind of 
variations w.r.t. $s_2$ and $s_1$, respectively. In other words, these observations certify that the conserved and nilpotent charges 
$Q$ and $\bar Q$  do {\it not} absolutely anticommute (i.e. $s_2 Q = -\,i\,\{Q, \bar Q\}\neq 0, s_1 \bar Q = -\,i\,\{\bar Q, Q\}\neq 0 $) as can be checked,
in a straightforward manner, by using the basic principles behind the relationships between the symmetry transformations  $s_1$ and $s_2$ and 
their generators (cf. ($B.5$)). The above observations are very important for our discussions on the  (anti-)chiral supervariable approach (i.e. ACSA) to our specific
${\cal N} = 2$ QM model of a 1$D$  SUSY harmonic oscillator.

We apply now the (anti-)chiral supervariable approach (ACSA) to derive the on-shell nilpotent 
(${s_1}^2 = {s_2}^2 = 0$) symmetry transformations $(B.2)$ for the ${\cal N} = 2$ QM model
of a 1$D$ harmonic oscillator. Towards this goal in mind, first of all, we derive $s_1$, for which, we generalize the 1$D$ basic variables 
$x(t), \psi (t), \bar\psi (t)$ to their counterparts supervariables $X(t, \bar \theta), \Psi (t, \bar \theta), \bar\Psi (t, \bar \theta)$ which are defined on the (1, 1)-dimensional 
anti-chiral super-submanifold of the general (1, 2)-dimensional supermanifold (on 
which our 1$D \,{\cal N} = 2$ SUSY QM model is generalized). The super expansions of the supervariables are
\[x(t)\longrightarrow  X(t, \bar\theta) = x(t) +\bar\theta\, f (t),\qquad \psi (t)\longrightarrow  \Psi (t, \bar\theta) = \psi (t) +\bar\theta\, (i\,b_1),\]
\[\bar\psi (t) \longrightarrow \bar\Psi (t, \bar\theta) = \bar\psi (t) +\bar\theta\, (i\,b_2),\eqno (B.8)\]
where  the secondary variables $(b_1, b_2)$ are bosonic  and $f$ is fermionic due to the fermionic $(\bar\theta^2 = 0)$ nature of $\bar\theta$.
In fact the (1, 1)-dimensional anti-chiral super-submanifold is parameterized by the superspace coordinates $Z^M = (t, \bar\theta)$. The following SUSY 
invariant quantities under $s_1$, namely,
\[ s_1 \psi = 0,\quad\qquad s_1 (x\, \psi) = 0,\quad\qquad s_1 (\dot x\, \dot\psi) = 0,\eqno (B.9)\]
lead to the derivation of the secondary variables $b_1 = 0$ and $f  = i\, \psi$ (see, e.g. [19] for details). In other words, we have the following expansions
\[ X^{(h_1)}(t, \bar\theta) = x(t) +\bar\theta\, (i\,\psi)\equiv x(t) + \bar\theta\, (s_1 x(t)),\]
\[\Psi ^{(h_1)} (t, \bar\theta) = \psi (t) +\bar\theta\, (0) \equiv \psi (t) +\bar\theta\, (s_1 \psi(t)),\eqno (B.10)\]
where the superscript ${(h_1)}$ denotes the supervariables defined  on the anti-chiral super-submanifold which have been derived  after the 
applications of the  SUSY invariant restrictions (cf. $(B.9)$). 
With the helps from the expansions $X^{(h_1)} (t, \bar\theta)$ and $\Psi^{(h_1)} (t, \bar\theta)$, we can compute the expression for $b_2 (t)$ in the expansion for $\bar\psi (t, \bar\theta)$. For this purpose, we note that the following quantity 
\[s_1 \Big[\frac {{\dot x}^2} {2} + i\,\bar\psi\,\dot\psi + \frac {\omega ^2 x^2}{2}\Big] = 0, \eqno (B.11) \] 
is on-shell (i.e. $\dot\psi  + i\,\omega\,\psi  = 0$) invariant. Hence, according to the basic tenets of ACSA, we demand the following
\[\frac {\dot X^{(h_1)} (t, \bar\theta)\;\dot X^{(h_1)} (t, \bar\theta)}{2} + i\,\Big[\bar\psi (t) + i\,\bar\theta\,b_2 (t)\Big]\,
\dot\Psi^{(h_1)}(t, \bar\theta) + 
\frac {\omega^2}{2}\,X^{(h_1)} (t, \bar\theta)\;X^{(h_1)} (t, \bar\theta)\]
\[ = \frac {{\dot x ^2 (t)}} {2} + i\,\bar\psi (t)\,\dot\psi (t)
 + \frac {\omega ^2}{2}\, x^2 (t),\eqno (B.12) \]
 which lead to the determination of $b_2 = i\,\dot x - \omega\,x$. Thus, we obtain the following super expansion for the supervariable
$\bar\Psi  (t, \bar\theta)$, namely;
 \[\bar\Psi ^{(h_1)} (t, \bar\theta)  = \bar\psi (t) + \bar\theta (-(\dot x + i\,\omega\,x)) \equiv \bar\psi (t) + \bar\theta (s_1 \bar\psi (t)).\eqno (B.13)\]
It is evident that we have derived the on-shell nilpotent transformation $s_1$ for the variable $\bar\psi (t)$ as the coefficient of $\bar\theta$
in the expansion $(B.13)$. It is self-evident that the superscript $(h_1)$ on the supervariable $\bar\Psi ^{(h_1)} (t, \bar\theta)$  denotes the expansion for this supervariable which has been obtained after the application of the on-shell SUSY invariant restriction $(B.11)$. A close and careful look at 
$(B.10)$ and $(B.13)$ demonstrates that we have derived {\it all} the symmetry transformation $s_1$ of $(B.2)$ as the coefficients of $\bar\theta$.
Hence, we also note that we have the mapping: $\partial_\theta\leftrightarrow s_1$.

In order to derive the other symmetry transformations $(s_2)$ of $(B.2)$ by using the ACSA to our ${\cal N} = 2$ SUSY QM model of 1$D$ harmonic oscillator, we generalize to basic 1D variables $x(t), \psi (t), \bar\psi (t)$ to the (1, 1)-dimensional {\it chiral} supervariables 
$X(t, \theta), \Psi (t, \theta), \bar\Psi (t, \theta)$  defined on the (1, 1)-dimensional {\it chiral} super-submanifold of the general (1, 2)-dimensional 
supermanifold, as 
\[ x (t) \longrightarrow X(t,  \theta) = x (t) + \theta\,\bar f (t), \quad \psi (t) \longrightarrow \Psi (t,  \theta) = \psi (t) + i\,\theta\,\bar b_{1} (t),\]
\[\bar\psi (t) \longrightarrow  \bar\Psi (t,  \theta) = \bar\psi (t) + i\,\theta\,\bar b_{2} (t),\eqno (B.14) \]
where $(\bar b_1, \bar b_2)$ are the secondary {\it bosonic} variables and $\bar f (t)$ is a {\it fermionic} secondary variable. These secondary variables are to be determined  in terms of the basic variables of the Lagrangian $(B.1)$ by using the on-shell SUSY invariant restrictions. In this context, we observe that 
$s_2 (\bar \psi) = 0,\;s_2 (x\,\bar \psi) = 0,\; s_2 (\dot x\,\dot{\bar \psi}) = 0$ imply the following restrictions 
\[\bar\Psi (t,  \theta) = \bar\psi (t), \quad X(t,  \theta)\,\bar\psi (t) = x (t)\, \bar\psi (t), \quad \dot X(t,  \theta)\,\dot {\bar\psi} (t) = \dot x (t)\,    \dot{\bar\psi} (t), \eqno (B.15)\]
which lead to $\bar b_2 (t)  = 0, \bar f (t) = i\,\bar\psi (t)$ (cf. [19] for details).
Thus, we have the following super expansions 
\[X^{(h_2)}(t,  \theta) = x (t) + \theta\,(-i\,\bar\psi (t))  \equiv x (t) + \theta\,(s_2 x (t)),\]
\[\bar\Psi ^{(h_2)} (t,  \theta) = \bar\psi (t) + \theta\,(0) \equiv \bar\psi (t) + \theta\,(s_2 \bar\psi),\eqno (B.16) \]
where the superscript $(h_2)$ on the supervariables, on the l.h.s., denote the expansions of the supervariables after the SUSY invariant restrictions $(B.15)$
and we note that we have already derived the transformations $s_2 x = -i\,\bar\psi,\; s_2 \bar\psi  = 0$ as the coefficients in the super expansions
$(B.16)$. This observation establishes that $\partial_\theta \, X^{(h_2)} (t, \theta) = s_2\, x (t)$ and 
 $\partial_\theta \,\bar\Psi^{(h_2)} (t, \theta) = s_2 \,\bar\psi (t)$
which, in turn, imply $\partial_\theta \leftrightarrow s_2$.

We have to now compute  the expression for $b_1 (t)$. In this context, we note that the following useful and interacting quantity 
is on-shell ($\dot{\bar\psi} - i\,\omega\,\bar\psi = 0$) SUSY invariant, namely;
\[s_2 \Big[\frac {{\dot x}^2} {2}  - i\,\dot{\bar\psi}\,\psi + \frac {\omega ^2 x^2}{2}\Big] = 0. \eqno (B.17) \]
The basic tenets of ACSA to ${\cal N} = 2$ SUSY QM system requires that the following SUSY restriction is true, namely:
\[\frac {\dot X^{(h_2)} (t, \theta)\;\dot X^{(h_2)} (t, \theta)}{2} - i\,\dot{\bar\Psi}^{(h_2)} (t, \theta)\,\Big[\psi (t) + i\,\theta \, b_1 (t)\Big] + 
\frac {\omega^2}{2}\,X^{(h_2)} (t, \bar\theta)\;X^{(h_2)} (t, \bar\theta)\]
\[ = \frac {{\dot x ^2 (t)}} {2} - i\,\dot{\bar\psi} (t)\,\psi (t)
 + \frac {\omega ^2}{2}\, x^2 (t),\eqno (B.18) \]
 where the expansions of $X^{(h_2)} (t, \theta)$ and $\bar \Psi^{(h_2)} (t, \theta)$ have been listed in $(B.16)$. The above restriction leads to 
 $b_1 (t)= -i\,(\dot x - i\,\omega\, x )$. Hence, we have the following super expansion:
 \[\Psi ^{(h_2)} (t, \theta)  = \psi (t) + \theta \,(\dot x - i\,\omega\,x) \equiv \psi (t) + \theta \,(s_2 \psi (t)).\eqno (B.19)\]
 It is evident that we have derived the transformation $s_2 \psi = (\dot x - i\,\omega\,x)$ as the coefficient of $\theta$ in the above expansion 
 where the superscript $(h_2)$ denotes that the supervariable $\Psi ^{(h_2)} (t, \theta)$ has been derived after the application of $(B.18)$.
 We note, once again, that we have: $\partial_\theta \leftrightarrow s_2$.

 Against the backdrop of our derivation of $s_1$ and $s_2$ from the applications of the on-shell SUSY invariant restrictions, we are in the position to expresses 
 the conserved charges $Q$ and $\bar Q$ in terms of the ACSA (to ${\cal N} = 2$ SUSY QM model of a 1$D$ harmonic oscillator) as follows    
 \[Q = \frac{\partial}{\partial\bar\theta}\Big[-i\,\bar \Psi^{(h_1)}(t, \bar\theta)\; \Psi^{(h_1)}(t, \bar\theta)\Big] 
\equiv \int d\bar\theta\,(-i\,\bar \Psi^{(h_1)}(t, \bar\theta)\;\Psi^{(h_1)}(t, \bar\theta)),\]
 \[~~~\bar Q = \frac{\partial}{\partial\theta}\Big[i\,\bar \Psi^{(h_2)}(t, \theta)\;\Psi^{(h_2)}(t, \theta)\Big] 
\equiv \int d\theta\,(i\,\bar \Psi^{(h_2)}(t, \theta)\;\Psi^{(h_2)}(t, \theta)),\eqno (B.20)\]
where {\it all} the symbols have been clarified earlier. It is straightforward to note that the nilpotency ($Q^2 = \bar Q^2 = 0$) of the charges $Q$ and $\bar Q$
\[\partial_{\bar\theta} Q = 0 \Longleftrightarrow s_1 Q = -i\,\{Q, Q\} = 0\Longrightarrow Q^2 = 0,\]
\[\partial_{\theta} \bar Q = 0 \Longleftrightarrow s_2 \bar Q = -i\,\{\bar Q, \bar Q\} = 0\Longrightarrow \bar Q ^2 =0,\eqno (B.21)\]
are deeply connected with the nilpotency $(\partial_{\bar\theta} ^2 = \partial_{\theta} = 0)$ of the translational generators $(\partial_{\bar\theta}, \partial_{\theta})$ along the  (1, 1)-dimensional (anti-)chiral super-submanifolds (of the general (1, 2)-dimensional supermanifold on which the ${\cal N} = 2$ SUSY QM model is generalized). The above observations, in turn, are very beautifully connected with the nilpotency $(s_1 {^2} = s_2 {^2} = 0)$ of the 
${\cal N} = 2$ SUSY symmetry transformations $(B.2)$. Thus, we observe that nilpotency properties of $(Q, \bar Q), (s_1, s_2)$ and 
$(\partial_{\bar\theta}, \partial_{\theta})$ are intertwined together in a meaningful and beautiful fashion.

We end this Appendix with the following concluding remarks which capture the key differences between the applications  of ACSA to BRST formalism
{\it and} ${\cal N} = 2$ SUSY QM models. First and foremost, we observe that there is {\it no} way to express (cf. ($B.20$)) the conserved and nilpotent 
${\cal N} = 2$ SUSY charges $Q$ and $\bar Q$ as the derivatives w.r.t. $\partial_{\theta}$ and  $\partial_{\bar\theta}$, respectively, which is {\it not}
the case with the (anti-)BRST charges $Q_{(a)b}$ that can be expressed as the derivatives w.r.t. {\it both} the translational generators 
$(\partial_{\bar\theta}, \partial_{\theta})$ along the (anti-)chiral super-submanifolds. Hence, within the framework of ACSA, the nilpotent 
and conserved ${\cal N} = 2$ SUSY charges $Q$ and $\bar Q$
do {\it not} absolutely anticommute (i.e. $Q\,\bar Q + \bar Q\, Q \neq 0$) {\it but} the nilpotent and conserved (anti-)BRST charges $Q_{(a)b}$
{\it absolutely} anticommute  (i.e. $Q\,\bar Q + \bar Q\, Q  =  0$). Finally, we also note that ACSA to BRST formalism is applicable to any arbitrary 
$D$-dimensional gauge as well as reparameterization invariant   theories {\it but} ACSA to ${\cal N} = 2$ SUSY QM models is applicable to 1$D$ systems {\it only}.

\end{document}